\documentclass[]{aastex631}

\usepackage{url}
\usepackage{bm}
\usepackage{amsfonts}
\usepackage{amsmath}
\usepackage{amssymb}

\newcommand{\be}{\begin{equation}}
\newcommand{\ee }{\end{equation}}
\newcommand{\dif}{\mathrm{d}}
\newcommand{\rkl}[1]{\left(#1\right)}
\newcommand{\ekl}[1]{\left[#1\right]}
\newcommand{\gkl}[1]{\left\{#1\right\}}

\newcommand{\pdfv}{\ensuremath{\mathrm{P}_V}}
\newcommand{\pdfm}{\ensuremath{\mathrm{P}_M}}

\begin{document}

\title{Planet formation regulated by galactic-scale interstellar turbulence}

\author[0000-0002-7501-9801]{Andrew J. Winter}
\affiliation{Universit{\'e} C{\^o}te d'Azur, Observatoire de la C{\^o}te d'Azur, CNRS, Laboratoire Lagrange, 06300 Nice, France}

\author[0000-0002-7695-7605]{Myriam Benisty}
\affiliation{Universit{\'e} C{\^o}te d'Azur, Observatoire de la C{\^o}te d'Azur, CNRS, Laboratoire Lagrange, 06300 Nice, France}
\affiliation{Universit\'{e} Grenoble Alpes, CNRS, IPAG, 38000 Grenoble, France}

\author[0000-0003-2253-2270]{Sean M. Andrews}
\affiliation{Center for Astrophysics | Harvard \& Smithsonian, 60 Garden Street, Cambridge, MA 02138, USA}



\begin{abstract}
     Planet formation occurs over a few Myr within protoplanetary discs of dust and gas, which are often assumed to evolve in isolation. However, extended gaseous structures have been uncovered around many protoplanetary discs, suggestive of late-stage in-fall from the interstellar medium (ISM). To quantify the prevalence of late-stage in-fall, we apply an excursion set formalism to track the local density and relative velocity of the ISM over the disc lifetime. We then combine the theoretical Bondi-Hoyle-Lyttleton (BHL) accretion rate with a simple disc evolution model, anchoring stellar accretion time-scales to observational constraints. Disc lifetimes, masses, stellar accretion rates and gaseous outer radii as a function of stellar mass and age are remarkably well-reproduced by our simple model that includes only ISM accretion. We estimate $20{-}70$ percent of discs may be mostly composed of material accreted in the most recent half of their lifetime, suggesting disc properties are not a direct test of isolated evolution models. {Our calculations indicate that BHL accretion can also supply sufficient energy to drive turbulence in the outer regions of protoplanetary discs with viscous $\alpha_\mathrm{SS} \sim 10^{-5}- 10^{-1}$, although we emphasise that angular momentum transport and particularly accretion onto the star may still be driven by internal processes.} Our simple approach can be easily applied to semi-analytic models. Our results represent a compelling case for regulation of planet formation by large-scale turbulence, with broad consequences for planet formation theory. This possibility urgently motivates deep observational surveys to confirm or refute our findings.
\end{abstract}



\section{Introduction} \label{sec:intro}

Planets form and begin their evolution within discs of gas and dust that persist for a few~Myr around young stars \citep[e.g.][]{Haisch01b}. The properties of these protoplanetary discs, and the physics that shapes them, is critical to our understanding of the origin of the (exo)planets we observe today. For many years, models for planet formation have been based on assumptions that the star-disc system by-and-large evolves in isolation from its surroundings. This assumption is now being challenged on a variety of fronts. External mechanisms that shape protoplanetary discs include irradiation from nearby massive stars \citep[][and references therein]{WinterHaworth22}, neighbour star-disc encounters \citep[][and references therein]{Cuello23} and the late stage in-fall of gas from the surroundings \citep[e.g.][]{Dullemond19, Kuffmeier20, Kuffmeier21, Kuffmeier23}. It is this last influence that is the topic of this work. 

Recent observations in molecular line emission \citep{Huang2021} and infrared scattered light \citep{Garufi2020, Benisty2023} are frequently revealing extended structures associated with protoplanetary discs, {on spatial scales several times the disc outer radius.} {In-fall has been suggested as a possible origin for these structures \citep{Dullemond19, Kuffmeier20, Kuffmeier23, Hanawa24}, which are highly complex, including large scale spirals and streamers around T Tauri \citep{Huang2023} and Herbig AeBe \citep{Boccaletti2020} stars, often associated with reflection nebulosity \citep{Gupta23}, accretion outbursts \citep{Seba2020,Hales2020,Dong2022} and misaligned inner discs \citep{Ginski2021}}. Evidence {that is suggestive of} in-fall events has been found not only during the earliest stages of disc evolution, when the star is still embedded in its natal environment \citep{Jorgensen09, Maury19, Codella24}, but also around stars such {as AB Aur \citep{Nakajima95, Grady99,Fukagawa04} DR Tau \citep{Mesa22} and SU Aur \citep{Ginski2021}, which are all mature star-disc systems of age $\sim 1{-}3$~Myr.} {A recent estimate from a sample size of $43$ disc-hosting stars suggests approximately $16$~percent are interacting with the ambient ISM \citep{Garufi24}. This material should fall onto the disc on a time-scale much shorter than the disc lifetime \citep{Dullemond19, Kuffmeier20}, suggesting that either these events occur much more frequently than observed or the external structure is replenished by the ISM for a prolonged period.} {Indeed, substantial mass accumulation at least over the first $\sim 1$~Myr of the disc life-time is hinted at by the fact that discs younger than $1$~Myr in $\rho$ Ophiucus and Corona Australis are somewhat more compact and lower mass than observed in $\sim 1{-}2$~Myr old regions \citep{Testi16, Testi22, Cieza19, Cazzoletti19, Williams19}. It is possible that discs contain a substantial fraction of recently accreted material for much of their lifetime, which would have significant consequences for disc evolution models.}

In-falling gas may be the consequence of Bondi-Hoyle-Lyttleton (BHL) accretion, where low velocity gas is captured due to the gravitational potential of the star \citep{Bondi52, Edgar04}. Theoretically, the rate of capture of material from the interstellar medium (ISM) via BHL accretion can be substantial while stars remain bound to a dense gas reservoir. {\citet{Padoan05} proposed this mechanism to explain the observed relationship between stellar accretion rates and stellar mass. This idea was supported by subsequent work \citep{Throop08, Klessen10}, while more recent analysis of turbulent hydrodynamic simulations in a periodic box have shown that late stage accretion may be significant source of mass and angular momentum for $\sim 1$~Myr old discs in high gas density environments \citep{Kuffmeier23, Pelkonen24, Padoan24}. In-fall has also been suggested as a mechanism to explain examples of substantial protoplanetary discs persisting around old stars \citep{Scicluna14, Derkink24}. However, assessing the importance of this process for the observed population of protoplanetary discs throughout their lifetime requires quantifying the temporal evolution of the turbulent ISM surrounding forming stars. This is challenging for hydrodynamic simulations that are both expensive and limited in integration time by the (turbulent) crossing time of the computational domain.}  

{In this work, we present a semi-analytic approach for tracking the BHL accretion rate over the disc lifetime. In a sense, this is a complement and extension of that of \citet{Throop08}, who tracked BHL accretion rates onto stars that are gravitationally bound to a dense gas reservoir.} Our aim here is to estimate the distribution of accretion histories over the whole disc lifetime for a population of young stars that resembles those surveyed for disc properties in the solar neighbourhood. This includes both the period in which the star is bound to the gaseous overdensity in which it forms \citep{Throop08} and once the bound gas has dispersed. {Statistically interpreting the role of BHL accretion on a disc population level requires simulating the turbulent cascade from galactic to protoplanetary disc size scales for the several Myr over which discs persist \citep{Klessen10, Colman22, Nusser22}.} To do so, we apply an excursion set formalism which is effectively zero dimensional \citep{Hopkins12}. This is therefore a highly efficient way of statistically following density fluctuations of the ISM, making it conducive to semi-analytic planet formation models \citep[e.g.][]{Weder23, Mordasini24}.

In the remainder of this work, we first present a model for tracking the local ISM conditions and resultant BHL accretion rate for stars evolving in a turbulent ISM (Section~\ref{sec:ism_props}). We then present a simple disc model and compare to observed disc properties in Section~\ref{sec:PPD_props}. We discuss our results in the broader context of planet formation in Section~\ref{sec:discussion} and draw conclusions in Section~\ref{sec:conclusions}.

\section{Modelling the Interstellar Medium}

\label{sec:ism_props}

\subsection{Overview}

In this section, our aim is to estimate the distribution of accretion histories for a population of  young stars that resembles those surveyed for disc properties in the solar neighbourhood. To do so, we must follow the velocity and density distribution of gas surrounding these young stars. This includes both the period in which {the star remains within the star forming region \citep[cf.][]{Throop08}, and once the bound giant molecular gas has dispersed. The latter, and to some degree the former, requires quantifying the properties of the supersonic, turbulent ISM. }

{Simulations of sustained, supersonic turbulent flows that are not self-gravitating have clearly shown that the volume-weighted probability density function for  density fluctuations is lognormal, with a variance that is a weak function of the Mach number \citep[e.g.][]{VazquezSemadeni94, Ostriker99}. A small fraction of the ISM in a steady state is therefore at a sufficient density to marginally overcome (turbulent) pressure support, and undergo gravitational collapse \citep{Evans99}. Regions which meet these criterion are known as giant molecular clouds (GMCs). These structures have long been known to obey so-called Larson relations, which broadly connect the size scale $R_\mathrm{cloud}$ and velocity dispersion $\sigma_v$ of molecular clouds in the galactic ISM \citep{Larson81}. Since original relations, improved data has established that generally the velocity dispersion $\sigma_v \propto R_\mathrm{cloud}^{1/2}$ \citep{Heyer04}, as expected from supersonic, rapidly cooling turbulence \citep{Burgers74}, although the normalisation of this relation may somewhat vary with environment \citep{Green24}. Any models aiming to track turbulent fluctuations in the ISM density and velocity, or the resultant properties of collapsing GMCs, must take into account these relationships.}

{To track turbulent fluctuations in the ISM, we adopt the excursion set approach described by \citet{Hopkins12}, with the key equations summarised in Appendix~\ref{app:excursion} for convenience. In brief, this approach involves quantifying the statistical distribution of densities or velocities at a given location in space as the sum of contributions to that density on larger scales (Appendix~\ref{app:excusion_maths}). A `trajectory' is defined as the cumulative sum of perturbations from the largest scale to the smallest scales (equation~\ref{eqn:trajectory}). These trajectories define the density and velocity on a single volume element, although the size of the volume element depends on the scale being considered.  For any given trajectory, we can determine whether the local density is sufficient to undergo gravitational collapse (Appendix~\ref{app:crit_density}), becoming a star forming region. Each scale also has an associated turbulent time-scale, based on which we can compute the temporal evolution of the turbulent fluctuations (Appendix~\ref{app:time_evol}). These latter two aspects allow us to estimate local density and velocity fluctuations during and after the star formation process.}

{In the remainder of this section, we explain the following steps in our procedure:}
\begin{enumerate}
    \item {\textit{Quantify the properties of local star forming regions (Section~\ref{sec:local_sfrs}):}} Apply the excursion set formalism {to establish a statistically representative sample of unstable trajectories, and therefore} determine the typical properties of star forming regions that are targeted for local disc surveys (within $\sim 200$~pc). 
    \item {\textit{Follow the collapse of molecular clouds and star formation (Section~\ref{sec:gmc_collapse}):}} {For a representative sample of star forming regions, with associated trajectories, we adopt a simple model for collapse and the star formation rate. This allows us to draw the formation times for a representative population of stars. Note that a star forming region may contain many stars, and is therefore not equivalent to the protostellar collapse or class 0 phase of stellar evolution.}
    \item {\textit{Track the dynamical evolution of stars in the bound cloud (Section~\ref{sec:bound_dyn}):}} For the representative stellar population, we track the dynamical evolution of stars based on the gravitational potential of the host star forming region.    
\item {\textit{Evolve temporal fluctuations of the turbulent medium (Section~\ref{sec:post_dispersal}):}} {Once the cloud is exhausted, the star is ejected into the turbulent medium. We continue to evolve the trajectory to track both density and velocity of the ISM during and after the collapse of the cloud according to equation~\ref{eq:Deltadelta_t}. }
    \item {\textit{Compute accretion rate onto the star-disc system (Section~\ref{sec:accrates}):}} Based on the local density and velocity during and after the collapse of the star forming region, we calculate the BHL accretion rate onto the star-disc system over time. 
\end{enumerate}
{In the remainder of this section, we describe these steps in more detail. For each stage, we also highlight caveats to our approach, which amount to points for development in future work.}

%We estimate the statistical evolution of the accretion rate semi-analytically by applying an excursion set formalism \citep{Hopkins12}. This method is effectively `0D', and is therefore a highly efficient way of statistically following density fluctuations of the ISM. 
\subsection{Properties of local star forming regions}
\label{sec:local_sfrs}
\begin{figure}
    \centering
    \includegraphics[width=0.5\columnwidth]{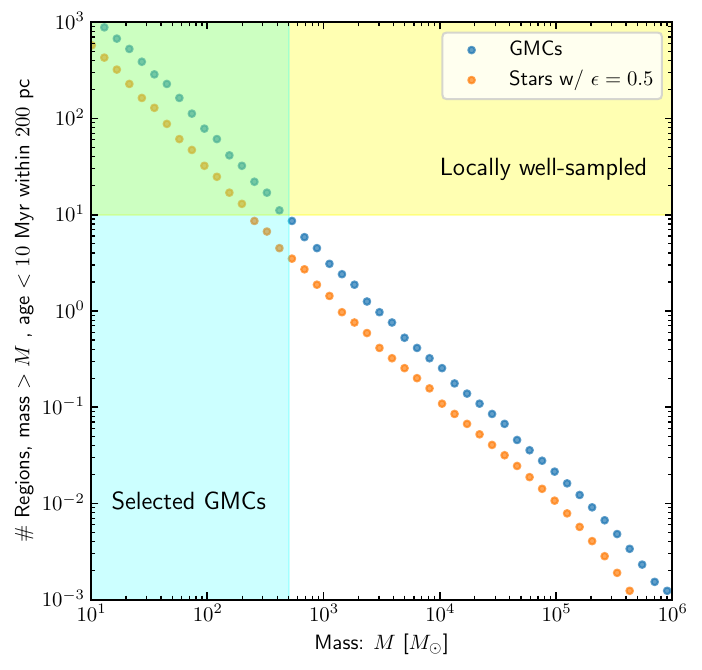}
    \caption{Sampling of the mass function via the excursion set formalism for GMCs (blue points) based on $10^5$ randomly drawn density perturbation trajectories in the solar neighbourhood. {Sampling errors are smaller than the marker size.} We have normalised the $y$-axis such that we show the average number of regions younger than $10$~Myr that are expected to be found within $200$~pc. We highlight in yellow the region which is well-sampled, corresponding to typical local star forming regions which are surveyed for protoplanetary disc properties. In cyan, we show the range of initial GMC masses we consider to generate our disc population synthesis. In orange, we show the same mass function assuming a fixed star formation efficiency $\epsilon=0.5$.}
    \label{fig:region_selection}
\end{figure}

\subsubsection{Approach summary}

{We first need to quantify the mass function of gravitationally unstable GMCs. To do so, we follow the approach of \citet[][]{Hopkins12}, as described in Section 3 of that work. In brief, this involves computing the probability of the turbulent medium collapsing at a given scale, based on the turbulent energy spectrum. We randomly sample a number of `trajectories', $N_\mathrm{traj}$, of which some number $N_\mathrm{coll}$ are unstable at some length scale $R$ corresponding to total mass $M$ (see Appendix~\ref{app:crit_density}). When weighted by the inverse of the relative volume that this star forming region represents, this yields an estimate for the mass function of unstable GMCs. However, to estimate the statistically sampled upper limit of the mass function among local star forming regions we must also estimate a normalisation constant.} The absolute frequency of star forming regions within a certain age range in the solar neighbourhood depends on the turbulent time-scale. This is itself a function of spatial scale:
\be
\tau_\mathrm{t} (R) = \eta\, t_{\rm cross} = \eta\,R/\langle v_{\rm t}^{2}(R) \rangle^{1/2},
\ee where $t_{\rm cross}$ is the crossing time-scale and $\eta \approx 1$ \citep{Pan10}. {The root mean square turbulent velocity $\langle v_{\rm t}^{2}(R) \rangle^{1/2}$ depends on the turbulent energy spectrum, as described in Appendix~\ref{app:excursion}.} An appropriate time-scale ${\tau}_\mathrm{t,0}$ over which the trajectory `resets' corresponds to the turbulent time-scale at given scale. {For simplicity, here we adopt} ${\tau}_\mathrm{t,0} \approx \tau_\mathrm{t}(h)$, reasoning that fluctuations at a scale $R\sim h$ are approximately the highest amplitude (we discuss this simplification in Section~\ref{sec:caveats_mfunc}). We can then estimate the differential number $\mathrm{d}N$ of star forming region younger than a certain age $\Delta \tau$ within a given volume $\Delta V$:
\begin{equation}
\label{eq:dNdM}
   \left\langle  \frac{\mathrm{d}N}{\mathrm{d}M} \right\rangle \approx \frac 1 {2^3} \frac{\Delta \tau}{\tau_\mathrm{t,0}} \frac{\Delta V \rho}{M} \cdot \frac{N_\mathrm{coll}}{N_\mathrm{traj}},
\end{equation}{where $\rho$ and $M$ are the density and mass of the star forming region respectively.} {We include a factor $1/2^3$ since given a region of radius $R_\mathrm{SFR}$, any other region with $R_\mathrm{SFR}$ is forbidden inside a distance of $2R_\mathrm{SFR}$ (such a region would merge with the original region).}

\subsubsection{Mass function}

We show the outcome of this calculation for $10^5$ independent trajectories in Figure~\ref{fig:region_selection}. We plot the number of star forming regions above a given mass $M$ within a spherical volume with radius {$R_\mathrm{obs} = 200$~pc ($\sim h$) younger than $10$~Myr.} This is approximately the radius within which the majority of protoplanetary disc statistics have been compiled \citep[for a review, see][]{Manara23} and the dispersal time-scale for protoplanetary discs \citep{Haisch01b, Pfalzner22}. We define the mass threshold below which we expect $>10$ star forming regions to be the threshold for being well sampled. This yields a mass limit $M_\mathrm{local} \approx 500 \, M_\odot$. Note, our definition of a star forming region is a region that is collapsing to form stars. This is in contrast to a broader complex that may not be collapsing globally. As an illustrative example, the Taurus-Auriga association has a total mass much larger than $500\, M_\odot$ \citep{Goldsmith08}. However, the complex as a whole is not collapsing to form stars. Instead collapse and star formation has occurred in several smaller sub-regions. It is these regions, or `NESTS' \citep{Joncour18}, that are the individual star forming regions in the sense appropriate to this work. In Taurus, these regions are typically a few tens of solar masses. The older region Upper Scorpius has a total stellar mass $\sim 1400\, M_\odot$ \citep{Preibisch02}, but is also composed of several individual populations \citep{Mire-Roig22b}. The nearest individual star forming region with total mass of several $10^3 \, M_\odot$ is the Orion Nebula cluster at a distance of $\sim 400$~pc \citep{Hil98, Reid09}. We conclude that, although our approach is approximate in that it neglects some physical processes that may influence local star formation, such as possible nearby supernovae \citep{Zucker22}, the upper mass limit we infer is reasonable. Our subsequent results do not depend strongly on this limit, as long as we exclude much more massive regions that may be dispersed prematurely by stellar feedback.

Motivated by this result, hereafter we we will truncate the GMC mass function above $500 \, M_\odot$. We will also exclude regions with masses $<10 M_\odot$, for which we expect the turbulent velocity to approach the sound speed ($\mathcal{M}\sim 1$). In principle, they may still be accurately sampled by the excursion set formalism if the density distribution does not diverge greatly from log-normal. However, we cannot choose an arbitrary small lower mass limit, since for much lower mass regions there is insufficient mass to form stellar mass objects. Experimentation shows that our results are not strongly dependent on our choice of lower mass limit {(see Section~\ref{sec:ll_sfr_caveats})}.

\subsubsection{Caveats for the mass function}
\label{sec:caveats_mfunc}
In the averaging for equation~\ref{eq:dNdM} we have ignored that similar mass unstable regions may be spatially correlated, since the probability of collapse on a given spatial scale is dependent on the density contribution from larger scales. However, \cite{Hopkins12} find that this bias is negligible for the low mass star forming regions, with $R_\mathrm{SFR} \ll h$. The shape of the mass function for the local, low mass GMCs relevant in the context of this work should therefore be correct. Clearly, star forming regions should still preferentially be found where the gas density is highest, while we expect using the average density is sufficient for $R_\mathrm{obs} \gtrsim h$. 

{We also highlight that our approach to computing the mass function of GMCs is not exact because we do not follow the time evolution of the interstellar medium -- i.e. we follow Section 3 of \citet{Hopkins12} rather than Section 7, which tracks turbulent fluctuations and assembly of merger trees. In principle, in order to compute the mass function of observed star forming regions we should take into account both the variations in the turbulent timescale at different spatial scales, and how the cloud evolves after it reaches critical density. To some degree, this can be folded in by evolving the turbulent medium via perturbations to each trajectory, as described in Appendix~\ref{app:time_evol}, and following Section 7 of \citealt{Hopkins12}. However, strictly this would also require self-consistently evolving the cloud, including collapse and growth. While this represents an aim for the future, we here wish to keep our model as simple as possible. We do not expect our simplification in assuming a single turbulent time-scale to strongly influence our results, primarily because the turbulent time-scale is a shallow function of $R$ ($\tau_\mathrm{t} \propto R^{1/2}$ for $R\ll h$), and the mass of the star forming region is a steep function of $R$ ($M\propto R^3$ for $R\ll h$), so that the turbulent time-scale is a very weak function of the mass of the star forming region ($\tau_\mathrm{t} \propto M^{1/6}$ for $R\ll h$). Indeed, we are vindicated in our assumptions in that we approximately recover $\langle \mathrm{d}N/\mathrm{d}M \rangle \propto M^{-2}$, as is found observationally across a wide range of galactic environments \citep[e.g.][]{Mok20}.}

\subsection{Collapse of the giant molecular cloud}
\label{sec:gmc_collapse}

\subsubsection{Approach summary}
In the early stage after a star forms, it occupies the local star forming region that is undergoing gravitational collapse. Since we are interested in the local density and velocity evolution throughout a star's lifetime, we require a simple model to follow the collapse of the GMC and formation of stars. To do so, {we first draw a representative sample of GMCs following Section~\ref{sec:local_sfrs}, each with an associated trajectory that we will subsequently use. For each of these regions, we wish to estimate how the region evolves during the collapse and star formation phase.} To do so, we adopt the approach of \citet{Girichidis14} to estimate the free-fall collapse and resultant star formation rate. The mathematics of the approach are described in Appendix~\ref{app:collapse}, but we briefly and qualitatively describe the model here. 

{We start with a gravitationally unstable GMC of radius $R_\mathrm{SFR}$, which is initially the largest scale on which a given turbulent trajectory is unstable (Appendix~\ref{app:crit_density}). This region collapses in free-fall. Physically, it does not do so monolithically, but because smaller sub-regions of the cloud may be higher density (shorter free-fall time-scales), they fragment first and continue to accrete from their parent structures \citep{Hoyle53}. By assuming that each part of the cloud is in free-fall, we can estimate the fraction of the cloud has collapsed beyond some density threshold to form stars from the initial probability density function for density on the scale $R_\mathrm{SFR}$. We define this density threshold by a free-fall time of $0.1$~Myr, although our results are not sensitive to this choice providing it is much shorter than the global collapse time-scale ($\sim 1{-}3$~Myr for our star forming regions -- see Figure~\ref{fig:density_SFR}). We emphasise that this does require knowledge of a geometrical density structure during the cloud collapse. We need only assume a geometry when when we compute the evolution of the local density and velocity for individual stars (Section~\ref{sec:bound_dyn}). We assume the collapse proceeds until the cloud is instantly dispersed when a star formation efficiency $\epsilon=0.5$ is reached, this value being commensurate with simulations of stellar core formation \citep{Matzner00}. }

\subsubsection{Caveats for the cloud collapse model}

Our philosophy in this work is to maintain a model that is as simple as possible, building a framework that can be developed in subsequent work. As a result, we have neglected numerous processes that may influence star formation. In particular, the cloud is exposed to heating by feedback processes from the stars as they form. A wide range of analytic and semi-analytic approaches have been developed to describe this process \citep[e.g.][]{Vazquez-Semadeni19}, and these may be adopted in future work. We are helped in this context in our interest in low mass star forming regions that are locally well sampled. We do not expect many massive stars or supernovae among these local regions, and we are therefore somewhat justified in neglecting feedback processes and exclusively considering gravitational collapse.

\subsection{Stellar dynamics in the bound cloud}
\label{sec:bound_dyn}
\subsubsection{Approach summary}
Having defined a local star formation rate {within an individual star forming region (which may form many stars),} we can draw formation times for an initial stellar population. Since these stars are initially bound to the cloud from which they form, we integrate the orbit of the star within the gravitational potential of the cloud, ignoring interactions with other stars. For numerical convenience we will calculate the gravitational potential with a smooth Plummer profile with a core radius $a=R_\mathrm{SFR}/2.6$, such that the outer radius of the star forming region is twice the half-mass radius. {The radius $R_\mathrm{SFR}$ evolves following free-fall collapse, as described in Section~\ref{sec:gmc_collapse} -- {i.e. $R_\mathrm{SFR}$ (and therefore $a$) evolves from its initial value following equation~\ref{eq:approx-radius-solution}, with $r_1$ being the initial size of the region.}} We place stars within the core radius, and we draw an isotropic velocity distribution with a dispersion assuming instantaneous virial equilibrium with the cloud. To track local density and relative velocity of the gas, we then assume the same smooth Plummer density profile \citep[as for][see discussion in Section~\ref{sec:caveats_bound_dyn}]{Throop08}. {We then evolve the star under the time-evolving potential, until the final dispersal of the cloud (when $M_* = \epsilon M_\mathrm{tot}$, for $\epsilon=0.5$). We do not modify the total mass in the region as gas is converted into stars, although given that we end the collapse phase at star formation efficiency $\epsilon=0.5$, this should only have an order unity influence on the total potential.}

\subsubsection{Caveats for accretion during the bound phase}
\label{sec:caveats_bound_dyn}
{During the bound phase, apart from our inclusion in the broader star forming context, this part of our approach makes similar simplifications as those of \citet{Throop08}. As did \citet{Throop08}, we neglect turbulent fluctuations of the medium within the collapsing region. Clearly gravitational collapse of the ISM modifies the local turbulent energy spectrum, and therefore a prescription for these fluctuations would not be straightforward. We expect our approach to recover typical accretion rates within a given region, but underestimates the variance in these rates. Unlike \citet{Throop08}, we consider the increase in overall density resulting from the collapse of the star forming region, an appropriate age dispersion of the stars, and a self-consistent time-scale in which stars remain in the bound region.}

{Another related simplification is that we also do not capture the protostellar/class 0 collapse phase, where for the first $\sim 10^5$~years the young system is clearly being fed by the environment \citep{Pineda23}. We simply form our star instantaneously at some time. In principle, a semi-analytic model that follows gravitational collapse and the influence on the local density distribution is possible \citep[e.g.][]{VazquezSemadeni24}. Such a treatment of the formation stage may be able to self-consistently capture both the initial mass function and produce a physically motivated initial disc from the residual angular momentum. This would again complicate our model, and we therefore do not present a full prescription for these early stages in this work. However, we explore the consequences of our assumed initial conditions for our disc evolution model in Section~\ref{sec:init_mass}. }

{Finally, we do not include the influence of encounters with other stars. This may be included in the future, although for the local star formation regions that are the topic of this work stellar densities are low. }

\subsection{Ambient density and velocity in the unbound ISM}
\label{sec:post_dispersal}

\subsubsection{Approach summary}
{Once the star forming region is dispersed, all the stars it hosts are instantaneously ejected into the background ISM. The background here is defined by the trajectory which evolves from the initial conditions determined at the moment of collapse of the star forming region. From this moment of collapse, we must then track the density and velocity perturbations at each scale in the trajectory following the procedure of \citet{Hopkins12}, as described in Appendix~\ref{app:time_evol}. At each time-step we (independently) perturb the velocity and logarithmic density perturbation at each scale stochastically, with a magnitude that depends on the turbulent time-scale. This density structure is not used to compute the ambient environment of the star until the collapsing stat forming region is dispersed.}

{We continue to evolve the ISM after the dispersal of the cloud and the star moves through the unbound ISM. The relative velocity of the star as it moves through the unbound ISM is the gas velocity at the relevant scale, minus the velocity of the star with respect to the bound cloud at the time of dispersal and the velocity of the cloud itself. The velocity of the cloud is fixed as the gas velocity at the time and scale of cloud collapse. Any star therefore has an associated ambient density and relative velocity at each scale.}

\subsubsection{Caveats for the turbulent fluctuations of the unbound ISM}

{The main simplifying assumption for this step is that local star formation does not influence the turbulent fluctuations of the ISM. To varying degrees, this is probably not the case. For example, a local increase in the gravitational potential should drive density enhancements to smaller scales, while stellar feedback drives density enhancements to larger scales. We discuss the role of feedback on the ultimate BHL accretion rates in Sections~\ref{sec:int_wind} and~\ref{sec:ext_wind}. We expect that our focus on lower mass regions at least somewhat mitigates the exclusion of these influences. Incorporating a prescription for how local star formation shapes the turbulent energy spectrum remains a goal for future work. }

\subsection{Accretion rate}
 \label{sec:accretionrate}
\label{sec:accrates}

\begin{figure}
    \centering
    \includegraphics[width=0.6\columnwidth]{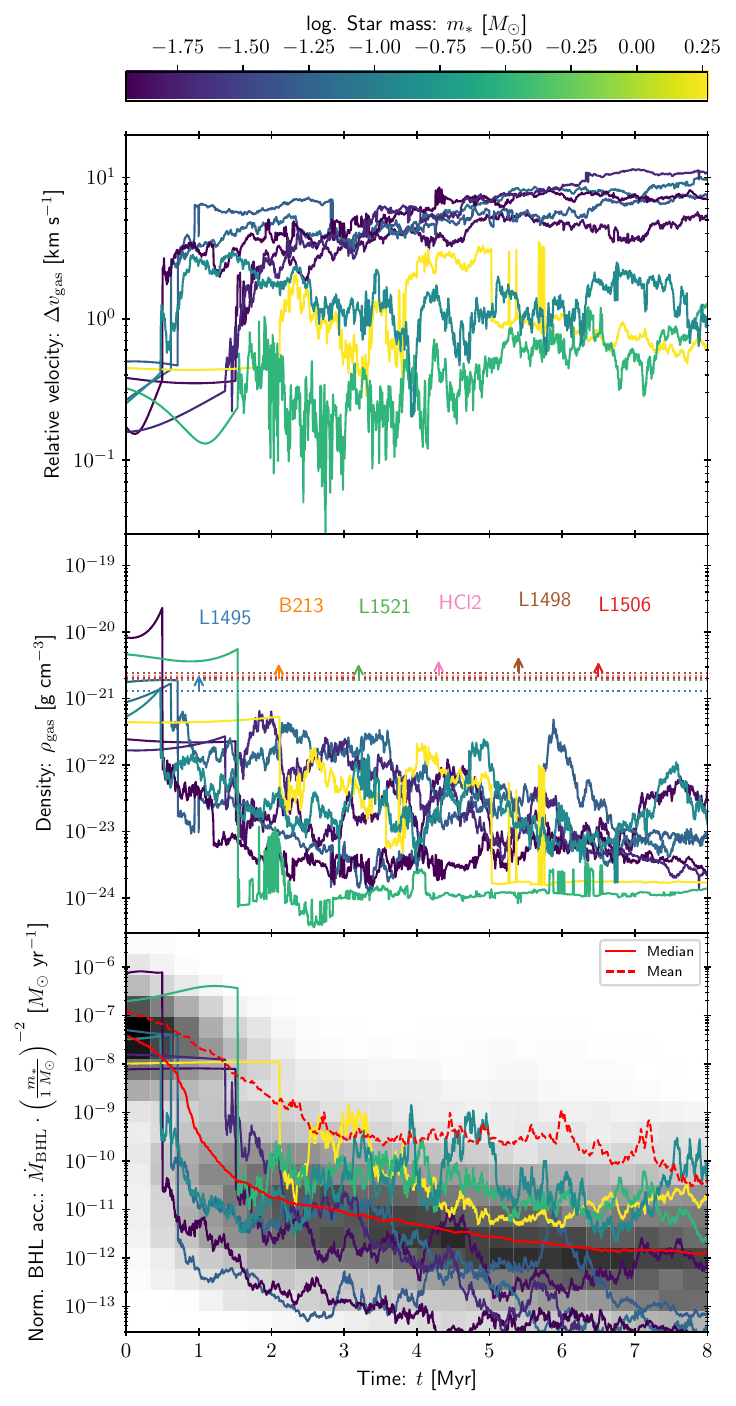}
    \caption{The relative velocity (top) and density (middle) of the local gas, for a subset of six stars in our sample of 500 trajectories. The BHL accretion rate (bottom) is normalised to be independent of mass $m_*$, by which lines are coloured. The gas density and velocity are shown at the accretion scale (see text for details). In the middle panel, the horizontal dotted lines are the mean gas density for some high density regions in Taurus \citep{Goldsmith08}. In the bottom panel, the red solid (dashed) line follows the median (mean) normalised BHL accretion rate for the entire sample, and a normalised two-dimensional histogram.}
    \label{fig:tevol}
\end{figure}
   
\subsubsection{Accretion radius }
BHL accretion has been the topic of many works \citep[for a review, see][]{Edgar04}. In the purest case, it is the gravitational focusing of ambient gas as the star passes through, resulting in a stagnation point in the wake of the star within which material is accreted. The instantaneous BHL radius within which material is captured is then defined:
\be
\label{eq:BHL_radius}
R_\mathrm{BHL} = \frac{2 G m_*}{(c_\mathrm{s}^2 +\Delta v_\mathrm{gas}^2)}.
\ee Streamlines with an impact parameter at the BHL radius $R_\mathrm{BHL}$ hit a stagnation point along the axis coincident with the relative upstream gas velocity vector. In the ideal case, in an homogeneous medium, they then accrete directly onto the star. In general, material within the cross-section carved out by the accretion radius $R_\mathrm{acc}$ (where typically $R_\mathrm{acc}=R_\mathrm{BHL}$) is captured by the star. At any turbulent scale, we can estimate the instantaneous rate at which a star captures material in such a scenario \citep{Bondi52, Shima85}:
\be
\label{eq:mdot_BHL}
    \dot{M}_\mathrm{BHL} \approx \pi R_\mathrm{acc}^2 \rho (\lambda c_\mathrm{s}^2 +\Delta {v}_\mathrm{gas}^2)^{1/2} \, , 
\ee where $\lambda = e^{3/2}/4 \approx 1.12$ in the isothermal limit. Material is accreted on the turbulent scale that maximises the accretion rate according to equation~\ref{eq:mdot_BHL}. We therefore define the ambient gas density $\rho_\mathrm{gas}$ and relative velocity $\Delta v_\mathrm{gas}$ to correspond to the quantities on this accretion scale. Before the dispersal of the formation cloud, the density and velocity are defined at the location of the star from the appropriate Plummer profile. 

{In some circumstances the radius within which capture can occur can be smaller than the BHL radius, $R_\mathrm{acc}<R_\mathrm{BHL}$. Broadly, some considerations that may alter the radius from which a star can accrete or suppress accretion entirely are as follows:}
\begin{enumerate}
    \item \textit{Tidal radius:} {The star is not isolated, but embedded in a larger scale gravitational potential. At the tidal radius $R_\mathrm{T}$, the tidal force of this larger scale structure balances the force induced by the stellar potential.} If the tidal radius $R_\mathrm{T}<R_\mathrm{BHL}$, then this radius limits the material that can be captured by the star. If a star is bound to a mass $M_\mathrm{b}$ at a radius $r_\mathrm{b}$, then:
\begin{equation}
    R_\mathrm{T} = r_\mathrm{b} \rkl{\frac{m_*}{3M_\mathrm{b}}}^{1/3}. 
\end{equation}Initially, this radius is determined by the collapsing cloud, where $r_\mathrm{b}$ is the instantaneous location of the star with respect to the centre of the region. Once the cloud has dispersed, we take the galactic values $r_\mathrm{b} = 8$~kpc and $M_\mathrm{b} = 10^{12} \, M_\odot$. 

    \item \textit{Turbulent scale:} Clearly the accretion radius cannot exceed the turbulent size scale under consideration, $R_\mathrm{t}$. 
    
    \item \textit{Photoevaporative wind:} {High energy photons from the central star can heat up the surface layers of the infalling column and drive a thermal wind, similar to the photoevaporative winds that are driven from the protoplanetary disc surface \citep[e.g.][]{Ercolano17}. }

    \item \textit{External irradiation:} {Stellar feedback from neighbouring stars on scales much larger than the disc may heat the ambient gas, increasing the sound speed and reducing the local density. }
\end{enumerate}

{In our fiducial model, we do not include photoevaporative suppression or large scale feedback, but discuss them in greater detail in Sections~\ref{sec:int_wind} and~\ref{sec:ext_wind}. Based on the above considerations, we define the accretion radius:}
\begin{equation}
    R_\mathrm{acc} = \operatorname{min}\gkl{R_\mathrm{BHL}, R_\mathrm{T}, R_\mathrm{t}}.
\end{equation}

\subsubsection{Accretion histories }

{In order to compute accretion histories, we draw a representative sample of stars from the representative sample of star forming regions. This involves drawing a star from a region with a probability that is proportional to both the total stellar mass of the star forming region, and at a time that is proportional to the instantaneous star formation rate within each region.} It is also now necessary to define the mass of a star to calculate accretion rates. Hereafter we draw stellar masses from a log-uniform distribution between $0.01-3 \, M_\odot$. This distribution is not important because we will resample when comparing to observations, except that we need to fully sample the stellar mass range relevant for drawing comparisons to protoplanetary disc surveys. {Although the stellar masses do not evolve self-consistently as the stars accrete in our model, the total mass that they would gain is nearly always a small fraction of the initial stellar mass.} We then draw 500 accretion histories from a representative sample of synthetic star forming regions, with a draw probability weighted by the stellar mass of the region. 

{We show the outcome for the evolution of the local density, relative velocity and BHL accretion rate (normalised to show the accretion rate assuming a solar mass star) for a subset of six stars in Figure~\ref{fig:tevol}. Given that density in particular can vary over several orders of magnitude, we also draw an approximate comparison to the typical densities in observed star forming regions. To do so, we adopt averaged density estimates from Table 4 of \citet{Goldsmith08} for a number of high density areas within the Taurus star forming complex. We convert the mass estimates $M_\mathrm{enc}$ within the quoted surface area $A_\mathrm{surf}$ to a density estimate by approximating spherical geometry, such that the volume averaged density is:  }
\begin{equation}
\label{eq:av_rho}
   \langle \rho_\mathrm{enc} \rangle_\mathrm{V} \sim \frac{3 M_\mathrm{enc}}{4\pi (A_\mathrm{surf}/\pi)^{3/2}}.
\end{equation}{However, this approximation probably yields an underestimate of the typical density within or close to collapsing star forming regions. This is because the area $A_\mathrm{surf}$ as defined by Goldsmith et al. is not limited to a single, collapsing star forming region. Instead, it contains both high and low density subregions, resulting in an underestimate of $\langle \rho_\mathrm{enc} \rangle_\mathrm{V}$ for the collapsing regions. Nonetheless, the typical $ \langle \rho_\mathrm{enc} \rangle_\mathrm{V} \sim 3\times 10^{-21}$~cm~g$^{-3}$ is commensurate with the high-end for the local gas densities implied by our model (see the middle panel of Figure~\ref{fig:tevol}). We conclude that the local densities for stars in our model are broadly consistent with the those observed in local star forming regions.}

The resultant BHL accretion rates, normalised for the a solar mass star, are shown in Figure~\ref{fig:tevol}, for a subset of six stars in our sample. The phase in which the star remains within the bound gaseous reservoir typically persists for $\sim 1$~Myr. During this phase, local low velocity and high density gas results in an early phase of enhanced BHL accretion at a median rate $\dot{M}_\mathrm{BHL, 1/2} \sim 10^{-8} \, M_\odot$~yr$^{-1}$ (Figure~\ref{fig:tevol}, bottom). This median rate then quickly drops to $\dot{M}_\mathrm{BHL,1/2} \sim 10^{-10} {-} 10^{-11} \, M_\odot$~yr$^{-1}$ in the age range $1{-}3$~Myr for a solar mass star. This is considerably lower than the typical rate of accretion onto the central star, $\dot{M}_\mathrm{acc} \sim 10^{-8} - 10^{-9} \, M_\odot$~yr$^{-1}$ \citep{Manara23}. One might therefore assume that BHL accretion no longer plays an important role for disc evolution. However, in a turbulent medium, vorticity of the in-falling material means that accretion is mediated via a (protoplanetary) disc \citep{Krumholz06, Kuffmeier20}. In this case, the change in accretion rate can be averaged over the time-scale $\tau_\mathrm{acc}$ at which angular momentum is transported within the disc. Empirically, this time-scale can be a considerable fraction of the disc lifetime \citep{Almendros-Abad24}. {In this case the median underestimates the true contribution of BHL accretion rate onto the disc to the accretion rate onto the star. By contrast, the mean accretion rate when averaged over all stars in our model is $\langle  \dot{M}_\mathrm{BHL} \rangle \sim 10^{-8} - 10^{-9} \, M_\odot$~yr$^{-1}$ at $1-3$~Myr, as shown as the dashed red line in Figure~\ref{fig:tevol}. However, the mean \textit{overestimates} the typical BHL accretion rate contribution for individual discs, since it is dominated by a small number of rapid accretors. Interpreting the contribution from BHL accretion to the stellar accretion rate for typical star-disc systems is therefore not a simple task. Compounding this problem further, discs that survive to comprise the observed sample may also preferentially be those which have been replenished by the ISM. Interpreting the role of BHL accretion for disc evolution therefore requires further calculation, which is the subject of Section~\ref{sec:PPD_props}.}

\section{Protoplanetary disc properties}
\label{sec:PPD_props}

\subsection{Simple disc model}
\label{sec:disc_evol}
In an homogeneous medium the velocities of steamlines are zero at the stagnation point. Captured material therefore has zero angular momentum and accretes directly onto the star. However, in an inhomogeneous medium, vorticity results in retention of a bound disc structure {that does not instantly accrete onto the star \citep{Krumholz06}}. For a given fluid element, we may then expect accretion onto the star to be mediated via a protoplanetary disc \citep{Moeckel09, Wijnen17}.

We now aim to generate a simplified `population synthesis' of discs shaped by BHL accretion. To do so, we wish to initially remain agnostic as to the driver of stellar accretion \citep{Lyn74, Tabone22}. We therefore consider a stellar accretion rate $\dot{M}_\mathrm{acc} = M_\mathrm{disc}/\tau_\mathrm{acc}$, for disc mass $M_\mathrm{disc}$ and fixed accretion time-scale $\tau_\mathrm{acc}$. For each of the discs in our sample discussed, we solve the initial value problem:
\be
\label{eq:ivp}
    \dot{M}_\mathrm{disc} =\dot{M}_\mathrm{BHL} -\dot{M}_\mathrm{acc}.
\ee 

{To evolve the disc mass, in our fiducial model we fix the initial $M_\mathrm{disc}(t=0) = 0$. This is equivalent to the assumption that the initial disc that forms due to core collapse is low mass, or rapidly depleted. This runs counter to the conventional assumption that the protoplanetary disc represents an evolved state of the primordial protostellar disc, which formed during the stellar core collapse. We make this decision as a `clean' test of our model; we wish to explore the properties of discs arising from BHL accretion alone. Such a model requires no additional free parameters. We emphasise that no specific observational constraints contradict a low initial disc mass. Any such evidence would need to connect the primordial circumstellar material, which forms over $\sim 10^2$~years, to the protoplanetary disc, which in the context of this work forms over $\sim 10^5$~years and evolves over $\sim 10^6$~years. {Although (massive) discs around Class 0 stars have been inferred \citep{Maury19, Tobin20} interpreting these observations in terms of the exact age of the (proto-)star (down to $\sim 10^4$~years), the nature of the disc and the time-scale over which any such a disc should survive is not simple. Since few, if any, unambiguous observations exist constraining the properties of primordial discs that are categorically around stars of age $\ll 10^5$~years, there is no reason to assume that such a primordial disc should be both long-lived and comparable in mass to the subsequently accreted one.} Indeed, it would be a remarkable coincidence if BHL accretion and protostar formation contribute an equal fraction of mass to the disc population. Nonetheless, we also discuss the impact of this assumption on our results in Section~\ref{sec:init_mass}.}

For $\tau_\mathrm{acc} = M_\mathrm{disc}/\dot{M}_\mathrm{acc}$ we assume a log-uniform distribution $\pm 1$~dex around $\log \tau_\mathrm{acc}/\mathrm{1 \, Myr} = -0.5$, based on visual inspection of the observed distribution \citep{Almendros-Abad24}. Empirically, the time-scale $\tau_\mathrm{acc}$ is not strongly correlated with stellar mass, but may be a complex function of the disc properties or stellar age. For example, visually interpreting Figure 12 of \citet{Almendros-Abad24}, the median accretion time-scale appears to be $\tau_\mathrm{acc,1/2} \sim 0.1$~Myr for the young region $\rho$~Ophiuchus, while $\tau_\mathrm{acc,1/2} \sim 1$~Myr for the older Lupus region. Indeed, we expect rapid BHL accretion to instigate turbulence in the disc, temporarily decreasing $\tau_\mathrm{acc}$. However, empirically it is not clear the degree to which any variation of $\tau_\mathrm{acc}$ is a survival bias, discs with shorter $\tau_\mathrm{acc}$ having been depleted {(we will show a similar evolution in our model which has fixed $\tau_\mathrm{acc}$ for each star in Section~\ref{sec:stacc_rates})}. While exploring time- or in-fall-dependent $\tau_\mathrm{acc}$ is an interesting avenue for future exploration, we initially maintain our simplified model, subsequently exploring the potential role of the ISM in driving a time-dependent stellar accretion rate (Section~\ref{sec:disc_turbulence}). 

Given the initial condition and accretion time-scale, we evolve each disc mass according to equation~\ref{eq:ivp} over the entire time period under consideration. However, we assume a simple criterion for disc dispersal: $M_\mathrm{disc}< 3\times 10^{-5}\,M_\odot$. We adopt this criterion when showing disc properties in subsequent analysis. This threshold is motivated empirically by the fact that few discs have masses below this value; our results are not strongly sensitive to this choice within a sensible range {(see Appendix~\ref{app:discdisp_dprops})}.

\subsection{Observational comparison} 

\begin{figure*}
    \centering
    \includegraphics[width=0.95\textwidth]{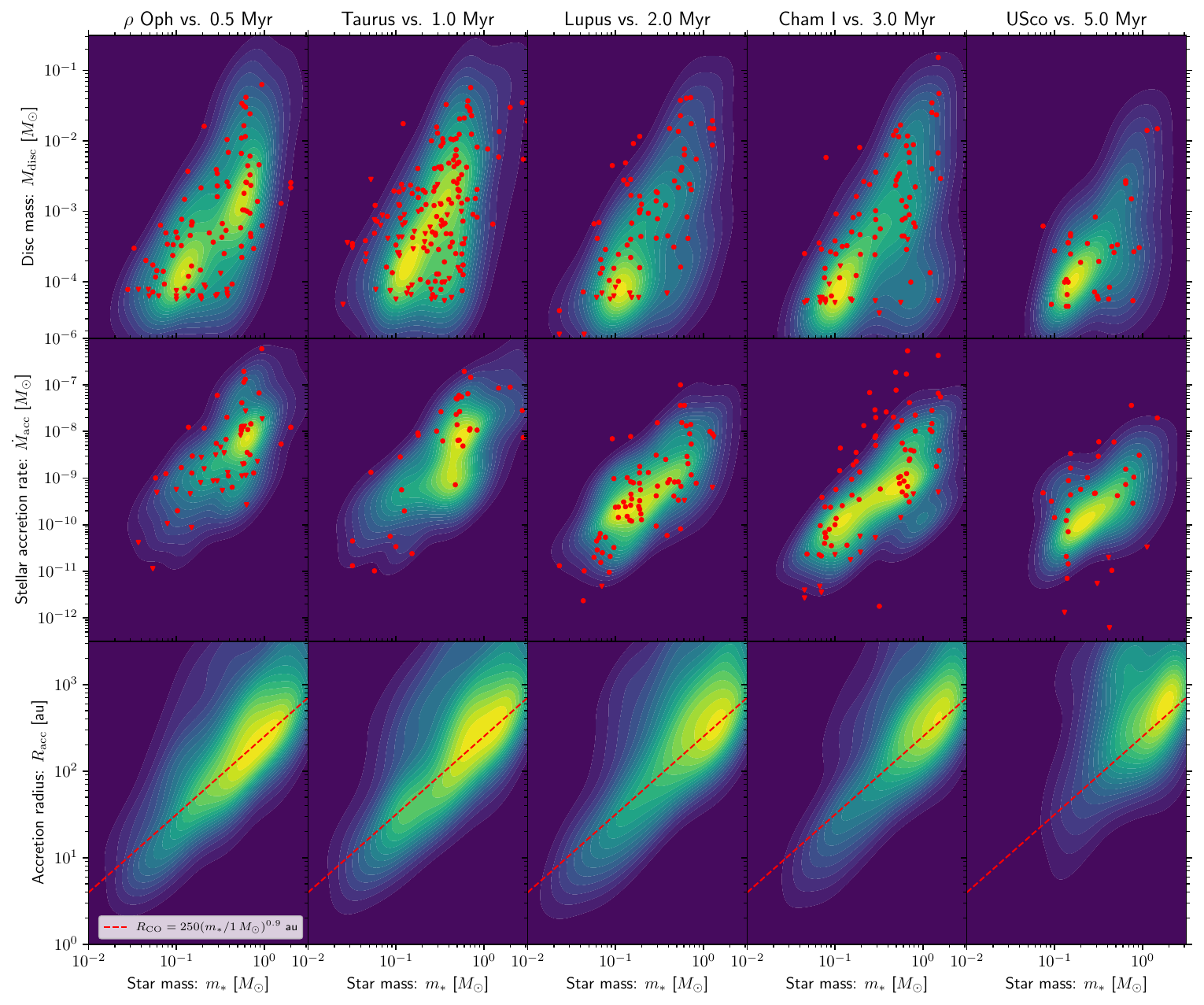}
    \caption{\label{fig:discprops}{
    Linear kernel density estimates (KDEs) for the disc mass (top), stellar accretion rate (middle) and accretion radii (bottom) versus stellar mass in our model compared to observations (red points, red dashed lines) in local star forming regions. We show results at times $0.5$, $1$, $2$ and $3$ and $5$~Myr. The red points represent the distribution of $100 \times$ dust disc masses (top), stellar accretion rate (middle) in $\rho$ Ophiuchus, Taurus, Lupus, Chameleon I and Upper Scorpius \citep[left to right --][]{Manara23}. When constructing the KDEs, model points are weighted to yield the same distribution along the stellar mass axis as in the observational samples, except in the bottom row. For comparison with the accretion radii, we show the empirically inferred relationship between CO outer disc radii as a red dashed line in the bottom panels \citep{Andrews20}.}}
\end{figure*}

We wish to compare our model to observed discs with comparable stellar ages. The ages of star forming regions can both be observationally uncertain \citep[e.g.][]{Soderblom14} and should exhibit physical scatter given the finite duration of the star formation process. However, nominally ages for the youngest populations of stars in the $\rho$ Ophiuchus complex are $\sim 0.1-1$~Myr \citep{Luhman99} in Taurus are $\sim 1{-}3$~Myr \citep{Krolikowski21}, in Lupus are $\lesssim 3$~Myr \citep{Galli_Lupus20} in Chameleon I have been estimated at $\sim 1{-}2$~Myr \citep{Galli21} and $\sim 3{-}6$~Myr \citep{Luhman07}, and Upper Sco are $\sim 5-10$~Myr \citep{Pecaut16, Luhman_USco20}. We compare these regions to our model {for ages of the individual stars (not the regions)}, $0.5$, $1$, $2$, $3$ and $5$~Myr respectively; these choices should be considered illustrative, not precise. For total disc masses we take $M_\mathrm{disc} = 10^2 M_\mathrm{dust}$, where $M_\mathrm{dust}$ is as quoted by \citet{Manara23}, inferred from the mm continuum flux with standard assumptions for opacity and temperature. In order to draw comparisons with each, we perform a Gaussian kernel density estimate with Scott's rule for bandwidth selection \citep{Scott92}. Where we compare to data directly, we weight the model data points to yield the same distribution in the stellar mass axis as for the observational dataset. In practice, this means dividing the stellar mass distribution into bins, and defining the weight of model data points in each bin $i$ to be $N_{\mathrm{obs},i}/N_{\mathrm{mod},i}$, where $N_{\mathrm{obs},i}$ and  $N_{\mathrm{model},i}$ are the number of observed and model points in bin $i$ respectively. The results of these comparisons are shown in Figure~\ref{fig:discprops}, in which we find remarkable agreement with the observed disc mass, stellar accretion rate, and outer disc radius distributions. We explore these findings in greater detail as follows.

\subsubsection{Disc masses} 

Disc masses are possibly the best metric of success of a disc evolution model, since they are time-integrated (i.e. not sensitive to temporal fluctuations) and relatively straightforwardly defined, albeit with some uncertain factors and observational caveats. Surveys inferring dust mass via mm flux are also relatively complete for several local star forming regions. In the top row of Figure~\ref{fig:discprops}, disc masses from our model show remarkable agreement with the observed data at all ages. Disc masses from our model fill-out the observed `wedge' in the $m_* -M_\mathrm{disc}$ plane.

\begin{figure}
    \centering
    \includegraphics[width=0.6\columnwidth]{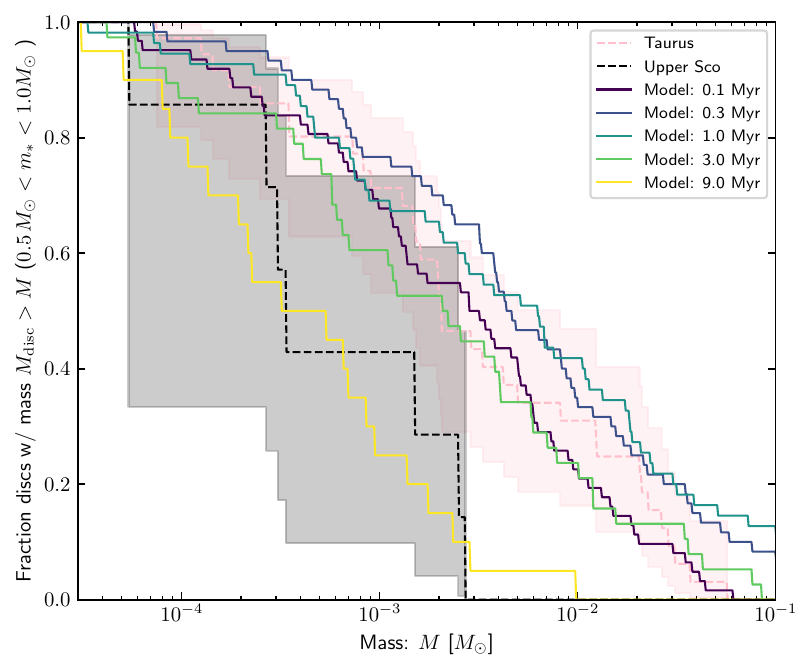}
    \caption{\label{fig:mdisc_cdf}Cumulative distribution of disc masses at different times in our model (solid lines) compared to the observed distribution (dashed lines) for Taurus (blue) and Upper Sco (orange). We include only two observed disc populations for clarify. Shaded region represents the statistical uncertainty interval for observational datasets.}
\end{figure}

We show the time evolution of the disc mass distribution in our model more quantitatively in Figure~\ref{fig:mdisc_cdf}, where we plot the cumulative distribution function for disc masses around stars $0.5\,M_\odot<m_* < 1\,M_\odot$. We also show observed distributions for Taurus and Upper Sco, leaving out the other regions for clarity. We can see that at $\sim 1$~Myr, the disc mass distribution in our model is very similar to the distribution in Taurus. This decreases after a few Myr until reaching a similar distribution to that seen in Upper Sco between $3$ and $9$~Myr. Within uncertainties, and given the caveat of a range of stellar ages in each region, our model appears fully consistent with the observed disc mass distributions. More speculatively, we note that the disc takes $\sim 0.3$~Myr to reach the maximum mass distribution. This is interesting in the context of $\rho$~Ophiuchus, which have discs that are more compact \citep{Testi16, Cieza19} and lower mass \citep{Williams19, Testi22} than $\sim 1{-}2$~Myr old regions. A similar result has been found in Corona Australis \citep{Cazzoletti19}. Our models suggest that this is plausibly the result of ongoing mass assembly proceeding from outside-in, via BHL accretion. 

\subsubsection{Stellar accretion rates}
\label{sec:stacc_rates}

The accretion rates from our model are entirely determined by the disc mass distribution, since we have imposed $\dot{M}_\mathrm{acc} \propto M_\mathrm{disc}/\tau_\mathrm{acc}$. They are therefore not an independent test of our model, but a validation or repudiation of this assumption. Given that we have anchored $\tau_\mathrm{acc}$ to observations, and we have found observational agreement with the disc mass distribution, as expected we also find excellent agreement for accretion rates for the majority of regions (middle row, Figure~\ref{fig:discprops}). {We note minor tension particularly in the case of Chameleon I, for which some accretion rates are somewhat larger than predicted by our model. This is plausibly the result of a younger population than the comparison age we have adopted. For example, \citet{Galli21} argue that the older esimate obtained by \citet{Luhman07} may be explained by systematic distance underestimate (then Chameleon I may be $\sim 1{-2}$~Myr old). Overall, we find excellent agreement between stellar accretion rates in our model and the observed samples. }

\begin{figure}
    \centering
    \includegraphics[width=0.6\columnwidth]{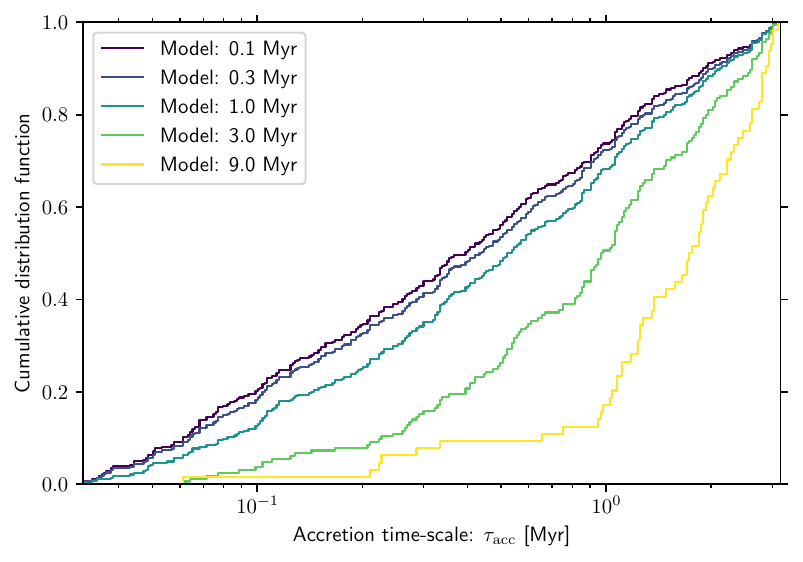}
    \caption{\label{fig:tau_acc}Cumulative distribution of the accretion time-scales $\tau_\mathrm{acc}$ at different times for surviving discs in our model. This evolution is purely due to discs with short accretion time-scales dropping out of the sample.}
\end{figure}

{We may also ask whether the data supports our assumption of a fixed $\tau_\mathrm{acc}$ distribution with stellar mass and time. It is possible, even probable, that this accretion time-scale does vary over time, particularly if internal disc turbulence is driven to some degree by the material inherited from the ISM (which is plausible, see Figure~\ref{fig:alpha_evol}). As previously discussed, empirically there is some variation in $\tau_\mathrm{acc}$ between regions \citep{Almendros-Abad24}. In this case, the true distribution of $\tau_\mathrm{acc}$ may be a function of the state of the ISM in each specific region.} {However, even for fixed $\tau_\mathrm{acc}$ for individual stars, the fact that star-disc systems with shorter $\tau_\mathrm{acc}$ lose their discs earlier mean that they drop out of the sample of surviving discs. For our model, this is shown in Figure~\ref{fig:tau_acc}, where we find that once discs are $\sim 3$~Myr old, the median $\tau_\mathrm{acc} \sim 1$~Myr, a factor three larger than for the initial disc population. This moderate increase is similar to the median for the observed distribution for discs of this age \citep{Almendros-Abad24}.} {We conclude that, while $\tau_\mathrm{acc}$ may vary over time for individual stars, such temporal variation is not required to produce the observed disc population.}

\subsubsection{Outer disc radii}
\label{sec:outer_radii}
We can also consider the distribution of gaseous disc radii $R_\mathrm{disc}$ we infer from our model. However, the appropriate definition of $R_\mathrm{disc}$ is less clear than for disc masses or stellar accretion rates {in terms of the correspondence between the model and observational data.} The observed outer radius in dust is dependent on variations in the dust opacity \citep{Rosotti19} and in gas molecular tracers, such as CO, outer radii are dependent on thermochemistry \citep{Trapman23}. Both are dependent on the definition of `outer radius', often defined by the radius that encloses some fraction of the total flux. {In our simple models, we do not explicitly define an outer radius. Simulations \citep{Krumholz06, Pelkonen24} and analytic estimates \citep{Padoan24} indicate that $R_\mathrm{disc} \sim R_\mathrm{acc} \propto m_*$ when {this disc is accreted from a turbulent medium}, with the disc radius typically a factor order unity smaller than the accretion radius. With the above caveats, we therefore attempt to compare $R_\mathrm{acc}$ with the observed outer disc radii. While there are fewer measurements of gaseous disc outer radii in local star forming regions, in the bottom row of Figure~\ref{fig:discprops} we compare our model to the empirically-inferred relationship for the outer gas disc radius as inferred from CO molecular line measurements: $R_\mathrm{CO} = 250 \left( m_* / 1\, M_\odot \right)^{0.9}$~au \citep{Andrews20} {typically a factor $\sim 2{-}3$ larger than the dust component, although with considerable scatter and some examples of much more extended CO radii \citep{Sanchis20}}. We again find a remarkable {similarity} between our model results and the observed disc properties, with a tail of large $R_\mathrm{acc}> R_\mathrm{CO}$. The agreement between the normalisation in our model and the data may be somewhat coincidental, since $^{12}$CO is usually optically thick so that $R_\mathrm{CO}$ probably overestimates where the bulk of the mass is, while $R_\mathrm{acc}$ also somewhat overestimates $R_\mathrm{disc}$. It is nonetheless clear that both the scaling and approximate normalisation of the $m_*-R_\mathrm{disc}$ relationship are predicted directly by BHL accretion.}

\subsubsection{Disc lifetimes}

\begin{figure}
    \centering  \includegraphics[width=0.6\textwidth]{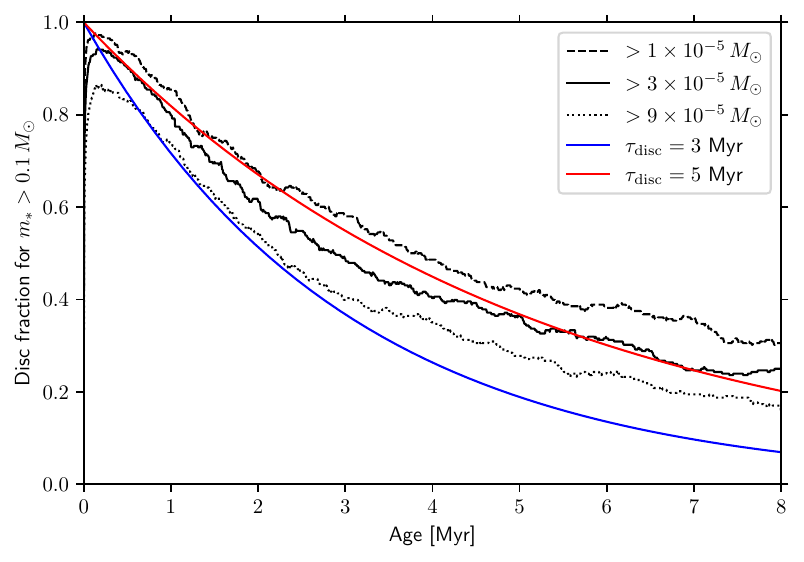}
\caption{\label{fig:disc_fraction}\small{Fraction of surviving discs in our model (solid black line), defined to be discs which have a mass $M_\mathrm{disc} > 3\times 10^{-5} \, M_\odot$ for stars with masses in the range $0.1{-}3 \, M_\odot$. {We also vary this threshold, shown by the dashed and dotted lines (factor $1/3$ and $3$ respectively).} We show the evolution of disc fraction given by $\exp\left( -t/\tau_\mathrm{disc}\right)$ for $\tau_\mathrm{disc}=3$~Myr \citep{Haisch01b} and $5$~Myr \citep{Ribas15} as blue and red lines respectively. }}
\end{figure}

For a fair comparison of disc property distributions, we must also confirm that the total fraction of discs included in our sample is similar in our model to those in star forming regions of a given age. We therefore show the disc fraction as a function of time in Figure~\ref{fig:disc_fraction}, by the definition $M_\mathrm{disc}>3\times 10^{-5} \,M_\odot$. We confirm that our model yields reasonable disc fractions over the range of times we consider in our disc property comparisons. We find a very similar disc fraction evolution to the exponential decay relationships suggested in the literature, typically with dispersal time-scale $\tau_\mathrm{disc} \sim 3{-}5$~Myr \citep{Haisch01b, Ribas15}. {In Figure~\ref{fig:disc_fraction}, we also show the disc fractions we obtain when multiplying the threshold mass by $1/3$ and $3$, finding only a moderate difference of $\sim 10$~percent between the most extreme choices. }

Interestingly, we find that {the disc fraction never reaches $100$~percent in our model}, but is maximal at $\sim 90$~percent. Observations indicate that, even for very young disc populations, the disc fraction never exceeds $\sim 80$~percent \citep{Michel21}. {However, when accounting for close binaries, this fraction could be closer to $100$~percent for single stars \citep{Kraus12}. The model prediction of a somewhat suppressed {early} disc fraction should therefore not be interpreted as a unique signature of BHL accretion, and may also depend on the presence or absence of a primordial disc.} We also find a somewhat flatter disc fraction as a function of time than a simple exponential decay with linear time exponent. Some authors to suggest such a flattened functional form for the disc fraction \citep{Pfalzner22}, supported by a non-negligible disc fractions in clusters that are tens of Myr old \citep{Galli21b}. Such surviving discs can be easily understood in the context of the stochastic accretion process that defines our model. 

We caveat this discussion in that we have not included any end-stage disc dispersal processes; in particular the photoevaporative wind driven by the central star \citep{Clarke01, Alexander06a, Gorti09, Owen10, Picogna19}. We would expect that, at least for very low mass discs, mass-loss may be dominated by such a wind. Our assumption is effectively that mass-loss in the wind is inefficient; in this case we can assume that the lifetime of an observable disc is essentially set by the accretion process in isolation. Once a disc is `dispersed' in our model, if at some later stage the disc replenishment occurs more rapidly than the true photoevaporation rate, then our model remains valid. Nonetheless, a future investigation of how photoevaporation may interact with BHL accretion for disc populations is warranted in future work.

\subsection{Replenished mass fraction} 
It is not clear immediately from Figure~\ref{fig:discprops} whether disc properties are set by a continuous process of BHL accretion. Does BHL just set the initial disc properties that then simply evolve over our imposed time-scale $\tau_\mathrm{acc}$, or is disc material constantly replenished over the disc lifetime? This is an important distinction because if disc material is constantly being replenished then, even with well-determined initial conditions, the distribution of disc properties can never be a direct probe of isolated disc evolution. 

To understand what fraction of material in observed discs should originate from `fresh' ISM material, we define a metric $f_{M, 1/2}$ to be the `half-life replenishment factor'. That is, the fraction of material within the disc that has been accreted within the most recent half of the disc life-time. In the absence of BHL accretion, equation~\ref{eq:ivp} has a trivial analytic solution that we label $M_\mathrm{isol} (t)$, given an initial mass. Then, if stellar accretion of the older material occurs before the accretion of younger material, and if stellar accretion is the only mechanism by which mass is removed from the disc, then at time $t_2$ the fraction of mass that has been added to the disc since time $t_1$  is 
\be
\label{eq:f_M}
f_{M} (t_1 | t_2, M_\mathrm{disc}(t_2)) = 1-M_\mathrm{isol}(t_1 | t_2, M_\mathrm{disc}(t_2)) /M_\mathrm{disc}(t_1). 
\ee From equation~\ref{eq:f_M} we define $f_{M, 1/2}$ such that $t_2 = t_1/2$. We expect this estimation of $f_{M, 1/2}$ to be a lower bound because we ignore any other disc dispersal processes. Processes such as planet formation or disc winds should be expected to reduce $M_\mathrm{isol}$, thus increasing $f_\mathrm{M}$.  

\begin{figure}
    \centering
    \includegraphics[width=\columnwidth]{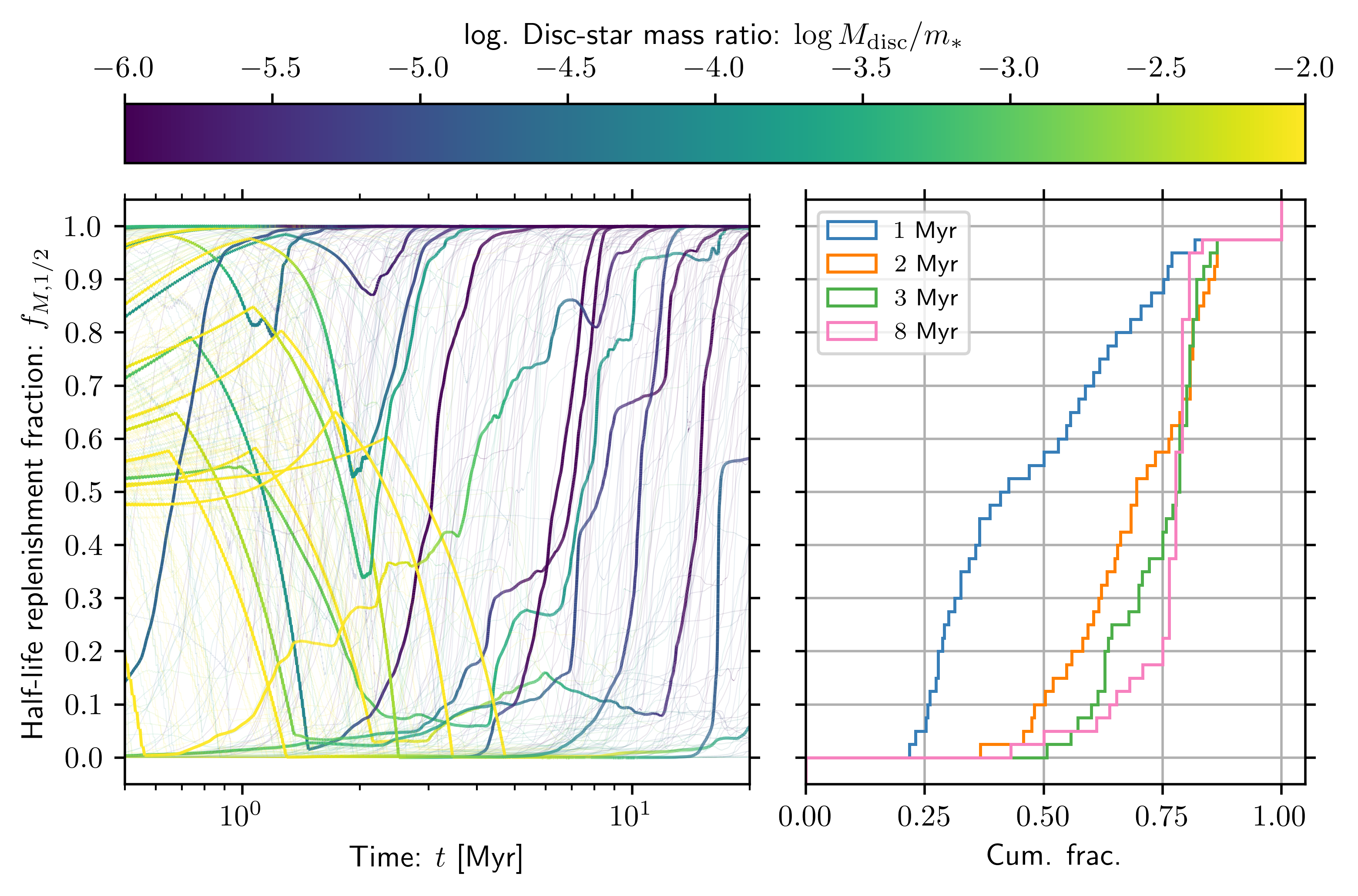}
    \caption{\label{fig:hlrf}The fraction of mass that has been accreted in the last half of the disc life-time, $f_{M, 1/2}$, as a function of time for $500$ discs in our model. {{On the left, we show the time evolution of $f_{M, 1/2}$ for all $500$ discs (faint lines), with a random subset of $10$ discs shown boldly for clarity.} These lines are coloured by} the ratio of the disc mass to stellar mass $M_\mathrm{disc}/m_*$, shown in the colour bar along the top. On the right hand side we show a cumulative fraction for $f_{M, 1/2}$ at $1$, $2$, $3$ and $5$~Myr. Only discs with a total mass $M_\mathrm{disc}>3 \times 10^{-5} \, M_\odot$ are included in this cumulative fraction, lower mass discs are assumed to be dispersed.  }
\end{figure}

Applying this metric in Figure~\ref{fig:hlrf}, we find a substantial fraction of discs composed of a high proportion of recently captured ISM material. {On the left panel we show how the replenishment fraction evolves over time for a random subset of discs. Typically, $f_\mathrm{M,1/2} \approx 0.5$ early on, then the fraction drops once the phase in which stars occupy the bound star forming region finishes. As the disc reduces in mass, discs often see an increase in  $f_\mathrm{M,1/2}$ later in their lifetime, when relatively modest BHL accretion rates can still provide substantial replenished mass compared to the existing disc mass. Not all discs follow this pattern however, with some discs having both high replenishment fractions  ($\gtrsim 0.5$) and considerable disc masses ($M_\mathrm{disc}/m_* \gtrsim 10^{-3}$) at relatively old ages ($\gtrsim 5$~Myr). The right panel shows that the fraction of discs composed of mostly replenished material drops from $\sim 70$~percent at 1 Myr, to $\sim 30$~percent at $2$~Myr and $\sim 20$~percent for discs older than $3$~Myr. This fraction therefore remains substantial throughout the disc lifetime.}

We conclude that disc properties are constantly `reset' for many discs in local star forming regions during their first few Myr of evolution. {{This would suggest that the distributions of observed disc properties should not be used as a test of theoretical models for isolated disc evolution.}} 

\subsection{Caveats and tests}

\subsubsection{Class 0 phase and the initial disc mass}
\label{sec:init_mass}

{
To simplify our model, we have assumed that a star initially forms instantaneously, without a period of growth that corresponds to the protostellar and class 0 phase. Indeed, for our fiducial model, we have made the assumption that the initial disc mass is negligible compared to the subsequently accreted material. As discussed, this assumption is justified in the desire for a `clean' test of the disc properties resulting from BHL accretion, retaining a simple model with few free parameters, and by the coincidence that an alternative would imply. However, given that this assumption is also unconventional, we can ask whether our conclusion regarding the degree of disc replenishment over time holds in the event of the coincidence that the primordial disc mass is comparable to the subsequently captured mass. }

{We explore this assumption in Appendix~\ref{app:initial_dmass}, repeating our experiment but assuming massive initial discs. While we still do not capture the class 0 disc assembly phase, this is equivalent to the opposite assumption as for our fiducial model; that the majority of the young discs mass is assembled within the first $\sim 10^5$~yr. From Figure~\ref{fig:hlrf_minit_0.01}, we conclude that the replenishment fraction is only influenced for young discs ($\lesssim 1$~Myr), after which the distributions remain similar (cf. Figure~\ref{fig:hlrf}). Our conclusion is that disc replenishment remains an important process throughout the disc lifetime therefore does not depend on this assumption.}

\subsubsection{Internal winds}
\label{sec:int_wind}

{As discussed in Section~\ref{sec:accrates}, photoevaporative winds driven by high energy photons from the accreting star may suppress the BHL accretion flow \citep[e.g.][]{EdgarClarke04}. While quantifying the degree to which BHL accretion is suppressed is a complex problem, and should be the subject of a dedicated study, in Appendix~\ref{app:feedback_BHL} we outline a simplified approach.}

{We discuss the results of including the internally driven wind in detail in Appendix~\ref{app:feedback_BHL}, but highlight two key conclusions as follows. The first is that photoevaporation leads to more discrete, episodic periods of accretion, particularly for high mass stars (Figure~\ref{fig:mdotevol_wind}). When stars are in a sufficiently low density environment, the wind completely suppresses accretion in our model, although in reality some accretion may still be possible. Secondly, the fraction of the disc mass that has been recently accreted is reduced by including the wind at late times (Figure~\ref{fig:hlrf__wind} compared to Figure~\ref{fig:hlrf}). {Despite including this wind,} our estimate suggests that $\sim 20$~percent of discs are {still} composed of mostly replenished material at an age of $2-3$~Myr. This fraction is not very strongly influenced because the periods where the majority of the mass is accumulated are the {rapid} BHL accretion phases where the dense accretion column is not disrupted by the wind. {Nonetheless, clearly it is plausible that internally driven winds can disrupt the BHL accretion flow, at least for stars $\gtrsim 2 \, M_\odot$ (Figure~\ref{fig:Rwind}), and therefore} this represents an important topic for future study.}

\subsubsection{External stellar feedback}
\label{sec:ext_wind}

{Apart from internal irradiation, we have excluded heating of the ambient medium by neighbouring stars. For older regions, with a lower ambient density, and more massive regions, with a larger number of OB stars, feedback should eventually become important for setting the ambient gas density and sound speed. This will in turn suppress BHL accretion, perhaps in conjunction with the internal wind discussed in Section~\ref{sec:int_wind}. Specifically, for feedback-heated regions, the sound speed and relative gas velocity would be enhanced, reducing $R_\mathrm{BHL}$ and possibly changing the mode of accretion from BHL to disc sweep-up (Section~\ref{sec:sweep-up}). While we do not expect this to strongly influence our results for the local regions which are the focus of this work (although perhaps more so in Upper Sco), this is clearly a factor that must be considered when interpreting disc properties in different environments \citep[e.g.][]{vanTerwisga23}.}

\subsubsection{Lower limit of star forming region mass}
\label{sec:ll_sfr_caveats}
{As shown in Figure~\ref{fig:density_SFR}, star forming regions of lower mass collapse over a shorter time-scale, with higher density. We might then expect the mass accumulation for these regions to be more dominated by this early phase. We have chosen a minimum mass of a collapsing GMC to be $10 \, M_\odot$, but if we reduce this we include more stars in lower mass regions that may then be more dominated by the initial stages of BHL accretion. However, we test the sensitivity of our results to this lower mass threshold in Appendix~\ref{app:lower_sfrmass_limit}, finding that in a statistical sense our results are not strongly sensitive to our choice.}

\subsubsection{Other caveats for the BHL accretion rate}

{There remain numerous caveats for our assumed BHL accretion rate. For example, complicating factors include magnetic fields, self-gravity of the gas, and a change in accretion mechanism when the BHL radius is within the disc radius (sweep-up). We discuss these issues in greater detail in Appendix~\ref{app:other_accretion}; generally we expect deviations of order unity from the nominal BHL accretion rate.  }

\subsection{Results summary}

{The qualitative results of our disc evolution calculations can be summarised concisely: \textit{BHL accretion is capable of reproducing observed disc masses, accretion rates, radii and lifetimes as a function of stellar mass and time. The BHL accretion rate is sufficient to have recently re-supplied a considerable fraction ($\sim 20{-}70$~percent) of discs of age $ 1{-}3$~Myr with a high fraction ($\gtrsim 50$~percent) of replenished material.}}

\section{Discussion}
\label{sec:discussion}
\subsection{Disc turbulence and stellar accretion}
\label{sec:disc_turbulence}

In our model, we have assumed that disc mass is accreted onto the star with a fixed time-scale $\tau_\mathrm{acc}$ via an undefined angular momentum transport process. This process could be internal to the star-disc system, driven by magnetic winds or by viscosity generated by (magneto-)hydrodynamic instability \citep{Tabone22, Lesur23}. {Here we assess whether BHL accretion, and late stage in-fall more generally, could drive turbulence and/or stellar accretion in protoplanetary discs. In terms of the latter, the case against BHL accretion as the origin for \textit{stellar} accretion (i.e. the accretion of the disc material onto the star) was concisely and influentially laid out by \citet{Hartmann06}, based on two arguments. We consider both of these arguments in Sections~\ref{sec:stell_acc} and~\ref{sec:low_gasdense}. }

\subsubsection{Angular momentum transport to the inner disc}
\label{sec:stell_acc}
{The first argument against BHL accretion as a driver of stellar accretion is that accreted material would presumably need to have negligible angular momentum in order to accrete close to the stellar radius. Both simulations of BHL accretion in a turbulent medium \citep{Krumholz06, Kuffmeier23, Pelkonen24} and the fact that disc radii far exceed the stellar radius imply that this is not the case. This would appear to demonstrate that BHL accretion cannot set stellar accretion rates. However, observationally it is clear that, at least for the highest accretors, (magnetically-driven) winds can be launched from radii $\lesssim 10$~au \citep[see Section 4.1 of][]{Pascucci23}, and other internal processes may also drive turbulence in the inner disc \citep[e.g.][]{Carr04, Najita09, Ilee14, Bosman23}. Then in-falling material only needs to reach down to these radii in order to eventually accrete onto the host star. Even if the initial angular momentum of material is significantly larger than this, some angular momentum transport in the outer disc may be driven by in-fall in two ways.}

{The first way is the misalignment of the angular momentum vector of material accreted at different times. For example, \citet{Kuffmeier24} recently showed that the angular momentum vector of accreted material in a turbulent medium is typically decorrelated on time-scales $\lesssim 1$~Myr. Clearly, adding fluid elements together with the same magnitude of angular momentum but offset angular momentum vectors will reduce the total specific angular momentum of the system. This can simply be understood as a delayed form of BHL directly onto the star. Such a process would presumably lead to shrinking of the disc over time, and may therefore leave a similar imprint in disc populations as MHD driven winds \citep{Tabone22, Manara23}. It would also result in misalignment between inner and outer discs, which is indeed evident in many discs \citep{Bohn22, Villenave24}. Whether or not this results in a continuous or bursty supply of material to the inner disc would depend on whether warps instigated by in-falling material decay on comparatively long time-scales \citep[e.g. the $\sim 10^5$~years estimated by][]{Kimmig24}, or whether associated instabilities drive high levels of turbulence that rapidly decays \citep{Deng20,Fairburn23}. }

{A second way in which BHL accretion may contribute to stellar accretion throughout the disc lifetime is via injecting turbulent energy directly into the disc. We can estimate how much turbulent energy can be supplied to the outer disc by the ISM during BHL accretion by computing the balance between the rate at which energy is deposited compared to the decay of that turbulent energy. The rate at which turbulent energy decays is \citep{Klessen10}:}
\begin{equation}
    \dot{E}_\mathrm{decay} = - \frac {3^{3/2}} 2 \frac{M_\mathrm{disc} v_\mathrm{t, disc}^3}{H_\mathrm{disc}},
\end{equation}where $H_\mathrm{disc}\sim 0.1 R_\mathrm{disc}$ is the disc scale height and the factor $3^{3/2}$ is included because $v_\mathrm{t, disc}$ is defined to be the turbulent velocity in a single dimension. The in-flowing turbulent energy from the ISM is:
\begin{equation}
     \dot{E}_\mathrm{in} = \frac 1 2 \dot{M}_\mathrm{BHL}v_\mathrm{in}^2.
\end{equation}If the velocity of the flow is dominated by the gravitational acceleration from the central star, then: \begin{equation}
    v_\mathrm{in}^2 \approx \frac{2G m_*}{R_\mathrm{disc}}.
\end{equation}Then, if the fraction of the energy that goes into turbulent motion in the disc is $\epsilon_\mathrm{t} = |\dot{E}_\mathrm{decay}/\dot{E}_\mathrm{in} | \leq 1 $ and the disc is in a steady state, then (rewriting equation 30 of \citealt{Klessen10}):
\begin{equation}
\label{eff_disk}
\left(\frac{v_\mathrm{t, disc}}{c_\mathrm{s}}\right)^3  \approx \frac{2\epsilon_{\rm{t}}}{3^{3/2}} \left(\frac{c_\mathrm{s}}{0.2 \,{\rm{km}\,\rm{s}^{-1}}} \right)^{-3} \left(\frac{\dot{M}_{\rm BHL}}{1.9 \times 10^{-8}\,M_\odot \,{\rm{yr}}^{-1}}\right) \left(\frac{M_{\rm disc}}{10^{-2} m_\star}\right)^{-1} \left(\frac{H_\mathrm{disc}}{R_\mathrm{disc}}\right).
\end{equation}Instantaneous steady state is probably a good approximation, because the decay timescale is \begin{equation}
    \tau_\mathrm{decay} = |E_\mathrm{t,disc}/\dot{E}_\mathrm{decay}| =H_\mathrm{disc}/v_{\mathrm{t,disc}}  \ll \tau_\mathrm{disc},
\end{equation}
where 
\begin{equation}
 {E}_\mathrm{t, disc} = \frac 1 2 M_\mathrm{disc} v_\mathrm{t, disc}^2.
\end{equation}Therefore, the BHL accretion rate $\dot{M}_\mathrm{BHL} $ needs to remain continuously sufficient to drive any observed turbulence throughout the disc lifetime. However, this requirement is probably met for most discs in the context of our model.

We can estimate the distribution of turbulent velocities statistically by directly computing the energy balance in our model. We can write:
\begin{equation}
    \dot{E}_\mathrm{t, disc} = \epsilon_\mathrm{t} \dot{E}_\mathrm{in} + \dot{E}_\mathrm{decay} +\dot{E}_\mathrm{acc},
\end{equation}where we include the accretion term:
\begin{equation}
    \dot{E}_\mathrm{acc} = -\frac 1 2 \dot{M}_{\mathrm{acc}} v_\mathrm{t,disc}^2,
\end{equation}although in practice $|\dot{E}_\mathrm{acc}| \ll |\dot{E}_\mathrm{decay}| $. Then we can solve the initial value problem, including the mass evolution of the disc as before. In principle, we could then compute the disc evolution fully consistently assuming only the ISM as the driver of internal physics, since we then have $\alpha_{\mathrm{SS}}$ or equivalently $\tau_\mathrm{acc}$. However, in practice, this requires introducing a free parameter $\epsilon_\mathrm{t} $, which may not be a constant. Strictly, it also means self-consistently evolving the disc surface density (or outer radius). For these reasons, self-consistently solving the disc evolution would considerably complicate our model. We therefore leave this to future work. Nonetheless, we can adopt sensible parameters: $\epsilon_\mathrm{t}=1$, $R_\mathrm{disc} = 250 (m_*/1\, M_\odot)$~au and $H_\mathrm{disc}=0.1 R_\mathrm{disc}$ with our fixed $\tau_\mathrm{acc}$ to obtain an estimate for the maximal turbulent disc velocity $v_\mathrm{t, disc}$ over time. 

{The outcome of this exercise is shown in Figure~\ref{fig:alpha_evol}, where we find a median $\alpha_\mathrm{SS} \sim 3\times 10^{-3}$ for $\sim 3$~Myr old discs. We can compare this to the $\tau_\mathrm{acc}$ distribution (Figure~\ref{fig:tau_acc}), assuming that:}
\begin{equation}
    \tau_\mathrm{acc} \approx 3.8  \left( \frac{\alpha_\mathrm{SS}}{10^{-3}}\right)^{-1} \, \left( \frac{H_\mathrm{s}/R_\mathrm{s}}{0.1}\right)^{-2} \, \left(\frac{m_*}{1\, M_\odot}\right)^{-1/2}  \, \left(\frac{R_\mathrm{s}}{50 \, \mathrm{au}}\right)^{3/2}\, \mathrm{Myr},
\end{equation}{where we define $H_\mathrm{s}$ as the scale height at the viscous scale radius $R_\mathrm{s}$, which is some fraction of the outer disc radius. Given the median $\tau_\mathrm{acc}$ in our model at $3$~Myr is approximately $1$~Myr, the median $\alpha_\mathrm{SS}$ is approximately sufficient to sustain accretion self-consistently in our model by direct kinetic energy injection alone (depending on disc geometry and injection efficiency).} In Figure~\ref{fig:alpha_evol} we also compare to observational constraints for turbulence in the outer regions of some discs. With the exception of DM Tau, which has turbulent velocity $v_{\rm{t,disc}} \sim 0.25{-}0.33 \, c_\mathrm{s}$, {and IM Lup which has $v_{\rm{t,disc}} \sim 0.4{-}0.6 \, c_\mathrm{s}$ \citep{PanequeCarreno24}  attempts to measure turbulence in protoplanetary discs have resulted in upper limits. For example, for the disc around HD 163296 this upper limit is $v_{\rm{t,disc}} \lesssim 0.05 \, c_\mathrm{s}$ \citep{Flaherty15, Flaherty17} and for TW Hydra it is $v_{\rm{t,disc}} \lesssim 0.08 \, c_\mathrm{s}$ \citep{Flaherty18, Teague18}. For $M_\mathrm{disc}/m_* = 0.01$, $H_\mathrm{disc}/R_\mathrm{disc} = 0.1$ and $\epsilon_{\rm{t}}=1$, then even substituting a modest $\dot{M}_{\rm BHL} \sim 10^{-11} \, M_\odot$~yr$^{-1}$ yields $v_\mathrm{t, disc} \sim 0.03 c_\mathrm{s}$, similar to the empirical upper limit on turbulence in some discs. This is approximately the median BHL accretion rate for a solar mass star in our model at an age of $\sim 3$~Myr (see the bottom panel of Figure~\ref{fig:tevol}). Typically values of $M_\mathrm{disc}$ are even smaller than $10^{-2} m_*$, so even greater turbulence may be imparted by the ISM. {Broadly, for discs older than $\sim 3$~Myr (as is the case for all those with upper limits on the turbulence) we expect relatively low levels of turbulence, while for younger discs, like DM Tau \citep{Zhang20} and IM Lup \citep{Mawet12}, turbulence driven by the ISM can also be larger. We conclude that sufficient turbulent energy is supplied to discs from the ISM to sustain the observational constraints on turbulence in the outer disc.

\begin{figure}
    \centering
    \includegraphics[width=0.7\textwidth]{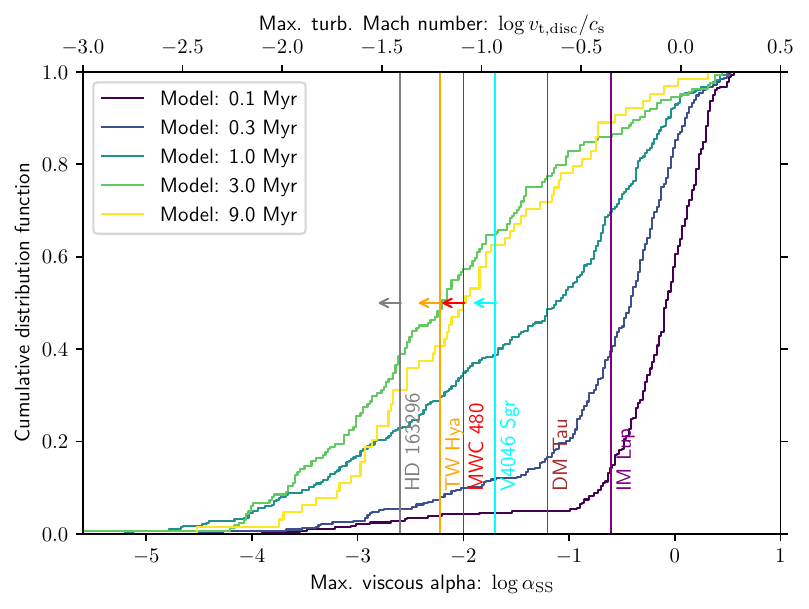}
    \caption{\label{fig:alpha_evol}Evolution of the cumulative distribution function for the turbulent Mach number (or viscous $\alpha_\mathrm{SS}$, top axis) for the discs in our model, assuming a maximal energy input efficiency $\epsilon_\mathrm{t}=1$. Each line colour represents a different disc age in our model, from $0.1$ to $9$~Myr. We show as coloured vertical lines some examples of discs with constraints on their turbulent velocity from the literature, several of which are upper limits \citep{Rosotti23}.}
\end{figure}

\subsubsection{Stellar accretion in regions of low gas density}
\label{sec:low_gasdense}
{The second objection raised by \citet{Hartmann06} is that stars in ionised HII regions still accrete. The typical density in such regions is far lower, and the sound speed and/or velocity dispersion much larger, such that the BHL accretion rate may be suppressed by several orders of magnitude. The example quoted by \citet{Hartmann06} are some stars in Trumpler 37 which were reported by \citet{SiciliaAguilar05} to be accreting substantially at a rate $\sim 10^{-8} \, M_\odot$~yr$^{-1}$. This is despite a sound speed $c_\mathrm{s,ext} \sim 10$~km~s$^{-1}$ (which we now distinguish from the sound speed in the disc $c_\mathrm{s,disc}$) and an electron density $N_\mathrm{e} \sim 3{-}15$~cm$^{-3}$ or $\rho_\mathrm{gas} \sim 2\times 10^{-23}$~g~cm$^{-3}$. The corresponding BHL accretion rate would therefore be $\dot{M}_\mathrm{BHL} \sim 10^{-13} \, M_\odot$~yr$^{-1}$ for a solar mass star, far too small to drive observed accretion rates. This would appear to rule out in-falling gas as the origin of angular momentum transport in at least these discs. However, this conclusion requires further investigation for the following reasons.}

{Firstly, as discussed above, turbulence injected into the outer disc may only serve to feed internally mediated transport in the inner disc. Therefore turning off the supply from the reservoir of material accreted from the ISM should not immediately end accretion onto the star. }

{Secondly, the small BHL radius resulting from large $c_\mathrm{s,ext}$ means that, at least initially after the gas is ionised, accretion onto the disc will be facilitated by sweep-up and not BHL accretion (see Appendix~\ref{sec:sweep-up}). If we re-calculate the sweep-up rate for a disc outer radius $250$~au (expected for a solar mass star), this increases the gas capture rate by two orders of magnitude to $ \dot{M}_\mathrm{su} \sim 10^{-11} \, M_\odot$~yr$^{-1}$. While this remains much smaller than the accretion rates reported by \citet{SiciliaAguilar05} in Trumpler 37, if the HII region is young \citep[$\lesssim 0.1- 1$~Myr --][]{Mellema06, Raycheva22}, then there may still be significant density variations that can lead to a spread in BHL capture rates. Indeed, \citet{SiciliaAguilar05} point out a population of embedded stars in Trumpler 37, which are evidence for at least some residual cold, dense gas.}

{Finally, as the mode of mass accretion changes, so does the rate of turbulent energy injection. This is both because the hot turbulent medium will typically have larger relative velocity $\Delta v_\mathrm{gas} \sim c_\mathrm{s, ext}$, and the material also injects thermal energy. In addition, if turbulent fluctuations are suppressed in the ionised medium, then accreted material contains little angular momentum and can fall directly into the inner disc regions. To estimate the resultant turbulent velocity within the disc, we can assume that $v_\mathrm{in} = \zeta c_\mathrm{s,ext}$, where $\zeta$ is a factor of order unity. Then:
\begin{equation}
     \dot{E}_\mathrm{in} = \frac 1 2 \dot{M}_\mathrm{su} 
 \zeta^2 c_\mathrm{s, ext}^2,
\end{equation}where $\dot{M}_\mathrm{su}$ is the sweep-up rate. The resultant steady state Mach number is:
\begin{equation}
    \frac{v_\mathrm{t,disc}}{c_\mathrm{s, disc}} \approx 0.016 \epsilon_\mathrm{t}^{1/3} \left( \frac{H_\mathrm{disc}}{10\, \mathrm{au}} \right)^{1/3} \left( \frac{M_\mathrm{disc}}{10^{-2}\, M_\odot} \right)^{-1/3} \left( \frac{\dot{M}_\mathrm{su}}{10^{-11}\, M_\odot \,\mathrm{yr}^{-1}}  \right)^{1/3} \left( \frac{\zeta c_\mathrm{s, ext}}{10\, \mathrm{km}\,\mathrm{s}^{-1}} \right)^{2/3} \left( \frac{c_\mathrm{s, disc}}{0.2\, \mathrm{km}\,\mathrm{s}^{-1}} \right)^{-1} 
\end{equation}While for our chosen normalisation yields $\alpha_\mathrm{SS} \sim 10^{-4}$, some combination of a factor three increase in the in-fall velocity or a factor eight in the sweep-up accretion rate would be sufficient to drive more substantial $\alpha_\mathrm{SS} \sim 10^{-3}$. Turbulence injected by warping could also still play a role in this context \citep{Deng20}. It is questionable whether either directly injected or warp-induced turbulence can sustain accretion into the inner disc for long in an ionised medium, but the moderate factors involved motivate further numerical experiment.}

\subsubsection{A case study: DM Tau}

We can also consider how the BHL accretion scenario fits with observations of individual discs. We consider the interesting scase of DM Tau, which is a star of mass $\sim 0.4{-}0.5\, M_\odot$ hosting a disc with a dust cavity and accreting at a rate $\sim 10^{-8}\, M_\odot$~yr$^{-1}$ \citep{Manara14, Francis20}. It has a turbulent velocity $v_{\rm{t,disc}} \sim 0.25{-}0.33 \, c_\mathrm{s}$, or viscous $\alpha_\mathrm{SS}\sim 0.1$ \citep{Flaherty20}. For this disc, a recent and relatively rapid period of accretion would be needed to drive the observed turbulence. Estimating $M_\mathrm{disc}/m_* \approx 0.03$ \citep{Andrews13} this would imply recent accretion from the ISM at a rate $\dot{M}_\mathrm{BHL} \gtrsim 2 \times 10^{-8} \, M_\odot$~yr$^{-1}$ that ended in the last $\sim 10^4$~years. This is similar to the current stellar accretion rate \citep{Manara14}, suggesting that the majority of the mass of the present day disc could have originated from this putative in-fall event. This scenario is plausible statistically within the context of our model. However, apart from the argument above, there is currently no particular evidence that DM Tau has undergone recent in-fall. This underlines the importance of a systematic search for observational signatures of late stage in-fall events.

\subsubsection{To what degree does in-fall mediate angular momentum transport?}

{In this section, we have considered whether in-fall can drive the separate but related phenomena of turbulence and stellar accretion in protoplanetary discs. It is clear that there is sufficient energy in the in-falling material to drive turbulence compatible with what is observed in the surface layers of the outer disc in the few cases where such constraints exist. This is not the same as the statement that this in-fall \textit{does} drive observed turbulence, to conclude as such will require careful case studies of observational examples. Similarly, understanding whether in-fall can regulate stellar accretion, either by angular momentum extraction or by turbulent energy injection, requires both more detailed theoretical and observational analysis. Probably internal processes must be responsible for angular momentum transport in at least some regions of the disc, but in-fall could play a role in regulating this process -- for example, by supplying material from the outer disc. We conclude that, although it is far from clear that these fundamental problems in disc evolution are solved by in-fall, ruling out its importance is not presently justified. }

\subsection{Appearance of BHL-accreting discs}

A pertinent question is what we expect the appearance of a disc undergoing BHL accretion to be. As previously discussed, when accreting material with with a high vorticity, the resultant disc outer radius can be of order the BHL radius \citep{Krumholz06, Kuffmeier23}. What is more uncertain is what should be the structure of a disc if the angular momentum of the accreted material is not aligned with the existing material. Depending on the relative angular momentum vector, this could result in `puffed up' discs, or misalignment between the inner and outer regions \citep{Kuffmeier21}. By contrast, at least the mm-dust component of discs is frequently observed to be extremely settled \citep{Duchene03, Pinte07, Pinte16, Villenave20, Villenave22, Pizzati23}. However, {(sub-)mm emission} is also commonly more centrally concentrated than the gaseous disc by a factor $\sim 2.5$ \citep{Ansdell18, Facchini19, Andrews20}, {which may be due to radial drift or optical effects \citep{Trapman19, Rosotti19}. It is not clear whether turbulence driven in the outer disc should be reflected in the dust distribution, that may have drifted inwards and/or settled vertically, even in the presence of the warping induced by in-fall \citep{Aly24}. In fact, we can consider one of the best known discs, HL Tau \citep{ALMA2015}, which has an extremely settled dust disc, evidenced by both dust height and polarisation that suggest low $\alpha_\mathrm{SS}\sim 10^{-5}$ \citep{Pinte16, Ueda21}. It is also one of the earliest known examples of a disc being fed by in-falling gas \citep{Hayashi93, Welch00, Yen19, Garufi2022} at a substantial rate \citep[$\gtrsim 5\times 10^{-9} \, M_\odot$~yr$^{-1}$ --][]{Gupta24}. This example therefore makes a compelling argument that the mm dust component of the disc is not necessarily strongly influenced by in-fall. It also suggests that turbulence may be substantially lower in the mid-plane with respect to the surface layers.} This represents an important topic for future study.

\subsection{Observability of in-fall events} 
\label{sec:observability}

{A valid  question to ask is the degree to which the model we present in this work predicts more abundant evidence of late-stage in-fall events from the data that is currently available. In fact, growing evidence suggests that a considerable fraction of discs are undergoing some kind of interaction with their environment. For example, from their recent scattered light survey of NIR polarimetric imaging for 43 discs in Taurus, \citet{Garufi24} reported that $16$~percent of those discs showed evidence of interaction with ambient material (although this sample is not necessarily representative). While inferring observational signatures of late stage in-fall requires full hydrodynamic and radiative transfer simulations \citep{Kieger24}, we can make a simple estimate for the density of gas we expect to be surrounding discs in our model. For accretion onto a disc from a radius $R_\mathrm{acc}$, the maximum free-fall time onto a star of mass $m_*$ is:
\begin{equation}
    \tau_\mathrm{ff, max} \approx 2 \times 10^3 \, \left(\frac{R_\mathrm{acc}}{250\,\mathrm{au}} \right)^{3/2}\left(\frac{m_*}{1\, M_\odot} \right)^{-1/2} \, \mathrm{years} .
\end{equation}Given a BHL accretion rate $\dot{M}_\mathrm{BHL}$, we can then estimate the column density of material external to the disc:
\begin{equation}
\label{eq:Next}
    N_\mathrm{ext} \sim  2\times 10^{19} \,   \epsilon_\mathrm{geo}^{-1}  \,  \left( \frac{\mu}{2.3}\right)^{-1} \left(\frac{\dot{M}_\mathrm{BHL}}{10^{-9} \, M_\odot\, \mathrm{yr}^{-1}} \right)^{-1/2}  \left(\frac{R_\mathrm{acc}}{250\,\mathrm{au}} \right)^{-1/2} \left(\frac{m_*}{1\,M_\odot} \right)^{-1/2} \, \mathrm{cm}^{-2}, 
\end{equation}where $\epsilon_\mathrm{geo}\leq 1$ is a geometric factor accounting for the non-uniform distribution of material and $\mu$ is the mean molecular mass. {We will assume $\epsilon_\mathrm{geo} = 1$ here, although based on the diverse ambient structures seen in observations, a wide range of $\epsilon_\mathrm{geo}$ may be plausible \citep[e.g.][]{Garufi24, Zurlo24}.} 

\begin{figure}
    \centering
    \includegraphics[width=0.6\columnwidth]{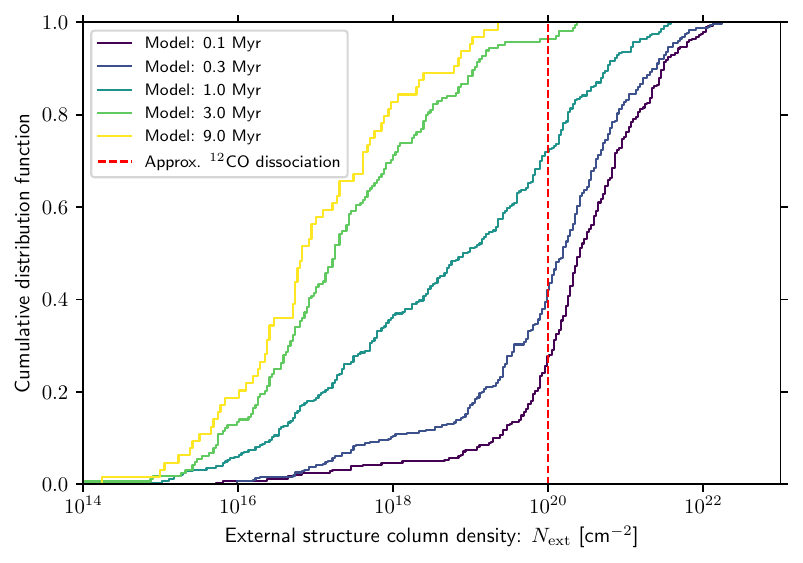}
    \caption{\label{fig:Next}The column number density of material assuming a mean molecular mass $\mu=2.3$ and $\epsilon_\mathrm{geo}=1$ following equation~\ref{eq:Next} for the surviving discs in our fiducial model. The colours of the lines represent the age of the discs. The red vertical dashed line represents an approximate column density for photodissociation of $^{12}$CO, based on a self-shielding column of $10^{15}$~cm$^{-2}$ \citep{vanDishoeck88} and abundance relative to hydrogen of $10^{-5}$ \citep{Trapman23}.}
\end{figure}

{The typical column density implied by equation~\ref{eq:Next} is relatively low, such that we would expect that in-fall is often challenging to trace observationally. We may consider the example of CO isotopologues, which are often the brightest tracers used to probe the gas  content of protoplanetary discs with ALMA. These isotopologues have a self-shielding column density of $\sim 10^{15}$~cm$^{-2}$ \citep{vanDishoeck88, Visser09}, and typical abundance relative to molecular hydrogen $\sim 10^{-5}$ \citep{Trapman23}}. {As an order of magnitude estimate, we may therefore expect column densities $\lesssim 10^{20}$~cm$^{-2}$ to have the CO photodissociated.} {From equation~\ref{eq:Next}, we plot the $N_\mathrm{ext}$ distribution over time resulting from the accretion rates in our model in Figure~\ref{fig:Next}. We see immediately in-falling CO would be photodissociated for the majority of discs in the age range $\sim 1{-}3$~Myr. Interestingly, in this age range, the fraction of discs above the threshold $N_\mathrm{ext} = 10^{20}$~cm$^{-2}$, approximately the CO self-shielding column, drops from approximately $30$~percent to close to $0$~percent. Given that this is the range of ages in Taurus, this would be commensurate with the inferred $\sim 16$~percent of discs with evidence of interaction with the ambient medium from scattered light observations \citep{Garufi24}. However, the physical column density down to which these observations are sensitive is not clear. While these estimates are approximate, they highlight the importance of deep surveys of the environments surrounding large numbers of discs. Such data could be applied to make statistical inferences about the prevalence of late stage in-fall, correlating it with disc properties and testing the theoretical predictions we have made in this work. }

\subsection{Implications}

Our results strongly suggest that disc properties are regulated by a continuous process of ISM accretion. This would have numerous consequences, many of which should be investigated in future work. We briefly summarise some examples as follows:
\begin{itemize}
    \item \textit{Disc sub-structures:} A diverse range of substructures have been observed in protoplanetary discs \citep{Andrews20}. We have demonstrated that the disc is undergoing constant perturbation by in-falling material. We would therefore expect some discs to exhibit structure due to this process \citep{Kuffmeier20}. For example, by accreting gas from the ISM with a misaligned angular momentum vector we expect disc warping \citep{Dullemond22, Kimmig24}, which may also produce observable shadows \citep{Marino15, Bohn22, Kieger24}. Both external structures and disc warping are evident in some discs \citep{Walsh17}, and it seems probable that they originate from late stage in-fall at least in some instances. 
    
    \item \textit{Accretion outbursts and short accretion times:} FU Orionis events are short accretion bursts that last $\sim 100$~years \citep{Hartmann96}, named for the archetypal example \citep{Herbig66, Herbig77}. {In-fall of material has been suggested as the origin of these outbursts \citep[e.g.][]{Vorobyov05}, alongside other mechanisms such as stellar encounters \citep{Bonnell92, Borchert22} and thermal instability \citep{Lin85, Clarke96, Lodato04}.} {It is unclear if in-fall events could produce short rise times, but may indirectly contribute by instigating instabilities in the disc \citep[e.g.][]{Deng20, Speedie23}.} Extended nebulosity surrounding several stars undergoing outburst supports the role of in-fall in at least some cases \citep{Zurlo24, Hales24}.  Outbursting star-disc systems may also be more broadly categorised as a subset of systems with short accretions time-scales \citep{Manara23, Almendros-Abad24}. Among such discs, a lifetime problem appears particularly pronounced for discs around some Herbig stars \citep{Grant23}. The assumptions of our model mean that we cannot directly explain these short accretion time-scales; we \textit{impose} this time-scale for our model. Nonetheless, given a fixed time-scale $\tau_\mathrm{acc}$, discs with short  $\tau_\mathrm{acc}$ may be episodically replenished in our model. Thus, to a degree the inferred accretion time-scale is decoupled from the expected frequency of observed discs, which no longer needs to be $\sim \tau_\mathrm{acc}/\tau_\mathrm{disc}$, for disc age $\tau_\mathrm{disc}$. Instead, the frequency of discs with a given $\tau_\mathrm{acc}$ would be dictated by the frequency of in-fall events that lead to rapid mass gain. {Indeed, in our model a small number of discs with very short accretion time-scales ($\sim 0.01$~Myr) are retained for $\sim 3$~Myr (Figure~\ref{fig:tau_acc}).} 
    
    \item \textit{Long-lived discs:} Stochastic replenishment of discs represents a simple explanation for so-called `Peter Pan' discs, which are (low mass) stars that are tens of Myr old but still exhibit a gas-rich accretion disc \citep{Silverberg20, Lee20, Galli21b}. It is not clear if Peter Pan discs are exclusively around low mass stars, or whether finding discs around M dwarf stars is just down to sampling of the mass function. Speculatively, this phenomenon may be related to the fact that individual massive discs appear across a number of regions of different ages, even where the majority of discs have been depleted \citep{Ansdell20}. Late stage in-fall seems a viable way to explain some of these long-lived discs. {This further motivates deep observational programs to characterise the super-disc scale gas surrounding protoplanetary discs.} 
    
    \item \textit{Planet properties:} Possibly the most exciting consequence of protoplanetary discs that are regulated by BHL accretion would be the potential consequences for planets. Multiple studies have shown that mature planet properties are dependent on stellar kinematics \citep{Hamer19, Winter20c}. Some of these differences could be related to covariance with stellar properties, such as age and metallicity \citep{Miyazaki23}. However, it is not clear that all correlations can be explained by stellar properties alone \citep{Longmore21, Winter21, Zink23}. If discs are constantly replenished by the ISM, stars that form at high velocity relative to the local rest frame should accrete less mass from the ISM after the dispersal of their formation cloud. This represents a direct theoretical link between planet formation and stellar kinematics. 
\end{itemize}

\section{Conclusions}
\label{sec:conclusions}
In this work, we have presented a novel method for computing the evolution of rate at which protoplanetary discs capture material from their surroundings via Bondi-Hoyle-Lyttleton (BHL) accretion. To this end, we have adapted the excursion set formalism developed by \citet{Hopkins12} to track the local density and velocity evolution of stars throughout the lifetime of the protoplanetary disc. Our three most important conclusions from this experiment are:
\begin{enumerate}
    \item Disc masses, radii and stellar accretion rates as a function of stellar mass and age can result directly from BHL accretion.    
    \item Material in the disc is constantly replenished over its lifetime. We estimate that $20{-}70$~percent of discs are mostly composed of material captured in the most recent half of their lifetime. This estimate is conservative in the sense that we include only stellar accretion, and ignore processes such as planet formation or winds that may further erode the existing disc. 
    \item  The energy influx due to in-falling material onto the disc is sufficient to drive empirically inferred protoplanetary disc turbulence. {If this in-fall produces turbulence in the outer disc, then we} expect a broad distribution for $\alpha_\mathrm{SS}$, with typical value $\alpha_\mathrm{SS}\sim 10^{-3}$, while some discs with much higher $\alpha_\mathrm{SS}\sim 0.1$ (as for DM Tau) and lower $\alpha_\mathrm{SS}\sim 10^{-5}$ are expected. {We highlight that stellar accretion is not necessarily driven by turbulence in the outer disc, and whether in-fall can contribute to global angular momentum transport to the inner disc remains an open question.}
\end{enumerate}
{Our approach is highly simplified, representing a novel method that can be built upon and adapted to semi-analytic models in order to include the role of BHL accretion in planet formation. Several processes may alter the rate of replenishment in the disc, but in particular stellar feedback may curtail late stage accretion for relatively massive stars ($\gtrsim 2\, M_\odot$) and stars born in regions of many OB stars. These feedback processes must be considered in future works.}

If the ISM {plays an important role in} disc evolution, this represents a considerable change of direction for protoplanetary disc theory, which has largely focused on processes operating on an isolated star-disc system. Consequences of this new picture of disc evolution are fundamental and wide-reaching, relating to every stage of planet formation. {The inextricable link between planet formation and the ISM underscores the importance of future observational and theoretical studies aiming to quantify how gas in-fall during formation influences protoplanetary disc structures, angular momentum transport and the observed exoplanet population. In particular, deep observational surveys aiming to correlate extended (in-falling) structures with disc properties and/or stellar accretion rates would test our findings, and are an urgent goal for the future.}

\section*{Acknowledgements} 

We thank the anonymous referee for an extremely thorough and thoughtful report, which helped us to considerably improve both the content and clarity of this work. We also thank Cathie Clarke, Alessandro Morbidelli, Sarah Jeffreson and Carlo Manara for useful discussions. AJW has received funding from the European Union’s Horizon 2020 research and innovation programme under the Marie Sk\l{}odowska-Curie grant agreement No 101104656. 
This project has received funding from the European Research Council (ERC) under the European Union’s Horizon 2020 research and innovation programme (PROTOPLANETS, grant agreement No. 101002188).

\bibliography{references}{}
\bibliographystyle{aasjournal}


\appendix

\section{Modelling turbulent fluctuations in the ISM}
\label{app:excursion}
\restartappendixnumbering

In order to quantify the mass distribution of star forming regions and the evolution of the ambient gas density and velocity, we apply an excursion set formalism as adapted by \citet{Hopkins12}, which was in turn adapted from the work of \citet{Bond91}. We briefly describe the approach of \citet{Hopkins12} here, using the same notation for clarity.

\subsection{Excursion set formalism}
\label{app:excusion_maths}
In the first instance, we note that for a region of space with mean density $\rho_{0} = \Sigma_0/2h$ (where we hereafter fix $\Sigma_0 = 12 \,M_\odot$~pc$^{-2}$) and Mach number $\mathcal{M} \gg 1$ on a spatial scale $R \sim 1/k$, the volume weighted distribution of gas density is well-described by a lognormal distribution: 
\begin{align}
\label{eqn:lognormal}
{\rm d}p(\delta\,|\,k) &= \frac{1}{\sigma_{k}\sqrt{2\pi}}\,
\exp \left( { - \frac{\delta^{2}}{2\,\sigma_{k}^{2}} }\right)\,
{\rm d}{\delta}  \\ 
\delta &\equiv \ln{\left(\frac{\rho}{\rho_{0}} \right)} - \left\langle{\ln{\left(\frac{\rho}{\rho_{0}} \right)}} \right\rangle
\end{align}
where 
\be
\left\langle{\ln{\left(\frac{\rho}{\rho_{0}} \right)}} \right\rangle  = -\frac{\sigma_{k}^{2}}{2}.
\ee  The density dispersion for $k$ can be estimated:
\begin{align}
\label{eqn:sigma.mach.general}
\sigma_{k} &\approx \left(\ln{\left[1 + \frac{3}{4}\,
\frac{\langle v_\mathrm{t}^2 (k) \rangle }{c_{s}^{2} + \kappa^{2}\,k^{-2}} 
\right]} \right)^{1/2},
\end{align}where we will assume throughout this work that the sound speed $c_\mathrm{s} = 0.2$~km~s$^{-1}$. Here $v_\mathrm{t}$ is the turbulent velocity, for which the average $\langle v_\mathrm{t}^2 (k) \rangle  \approx \sigma_\mathrm{g}^2 (k) \sim k E(k) \propto k^{1-p}$ where $E(k)$ the turbulent energy spectrum. Throughout this work we will fix $p=2$, appropriate for supersonic, rapidly cooling turbulence \citep{Burgers74}. On the smallest scales we modify $\langle v_\mathrm{t}^2 (k) \rangle^{1/2} $ to be the sum in quadrature of the power-spectrum component and the sound speed due to the local thermal pressure. The second term in the logarithm in equation~\ref{eqn:sigma.mach.general} is a modification of the normal term, which would be proportional the square of Mach number. The additional term in the denominator is proportional to the square of the galactic epicyclic frequency $\kappa$, for which we will assume a flat rotation curve with $\kappa = \sqrt{2}\Omega$, adopting the orbital frequency $\Omega = 2.6\times 10^{-2}$~Myr$^{-1}$ appropriate for the solar neighbourhood. This modification is necessary because at large $R$ the differential rotation $\kappa/k$ plays a similar role to thermal pressure on small scales. 

Assuming that the turbulent properties of the medium depend only on local gas properties at that scale, then the distribution of densities averaged over any spatial scale is also log-normal with variance:
\be
\label{eqn:variance.R}
\sigma^{2}(R) = \int {\rm d}\ln{(k)}\,\sigma_{k}^{2}(k)\,|\tilde{W}(k,\,R)|^{2}
\ee 
where $\tilde{W}(k,\,R)$ is the Fourier transform of the real-space window function $W(x,\,R)$ defined to include the relevant range of spatial scales. Clearly, the density distribution on progressively smaller spatial scales depends on the contribution of all larger scales. Following \citet{Hopkins12}, we define $\tilde{W}(k,\,R)$ to be a Fourier-space top-hat function: $\tilde{W}(k\,|\,R) = 1$ if $k\le R^{-1}$ and 
$\tilde{W}(k\,|\,R) = 0$ if $k> R^{-1}$. 

With these quantities defined, the density field at some infinitesimal volume can be understood as a random walk through Fourier space, which is simple to quantify for a Gaussian random field with the Fourier top-hat window \citep{Bond91}. In this case, the probability of a change in logarithmic density perturbation from $\delta_{1}$ to $\delta_{2}\equiv \delta_{1}+\Delta \delta$ 
given a change from scale 
$k_{1}$ to $k_{2}$ is given by 
\begin{align}
\label{eq:pdf_Deltadelta}
p(\delta_{1}+\Delta\,\delta)\,{\rm d}\,\Delta\,\delta &= 
\frac{1}{\sqrt{2\pi\,\Delta\,S}}\,\exp{{\Bigl(}-\frac{(\Delta\,\delta)^{2}}{2\,\Delta\,S}{\Bigr)}}
{\rm d}(\Delta\,\delta) \\ 
\Delta\,S &\equiv S_{2}-S_{1} \equiv \sigma^{2}(R_{2})-\sigma^{2}(R_{1})
\end{align}
where we define the variance
\be
S(R) \equiv \sigma^{2}(R).
\ee With this result, we can therefore define a `trajectory' for the variation of density across a range of discretely defined spatial scales. Starting on some sufficiently large initial scale $R\rightarrow \infty$, we have $\delta \rightarrow 0$ and $\sigma(R) \rightarrow 0$. We then have the well-defined initial conditions for the walk down through scales starting at some large $R_\mathrm{max}$, with increments
in $R$, $\Delta\,R_{i}$. A single trajectory is then:
\be
\label{eqn:trajectory}
\delta(R_{i}) \equiv \sum_{j}^{R_{j}>R_{i}} \Delta\,\delta_{j},
\ee where each $\Delta\,\delta_{j}$ is drawn from the probability density function defined by equation~\ref{eq:pdf_Deltadelta}. We will hereafter define a grid with a maximum scale $R_\mathrm{max} = 5h$ down to a minimum scale $R_\mathrm{min} < 0.01$~pc, each {successive turbulent scale in the trajectory being $95$~percent of the scale above. As shown in Figure~\ref{fig:deltaS_S}, choosing this value ensures that the fractional contribution to the overall variance at each successive spatial scale remains small ($\Delta S/S<0.1$). In other words, any fluctuations in density or velocity on scales between the those for which we calculate the structure of the medium will not contribute significantly to the local field. If instead we use $\Delta R/R \lesssim 0.5$, this is no longer true; we may miss scales in our calculation that dominate the local density or velocity on an intermediate scale for $R\sim h$ (see red lines in Figure~\ref{fig:deltaS_S}).} 

{We have defined $S(R)$ as the variance in the logarithmic density, but the same argument applies to any property which is Gaussian distributed with a known contribution to the variance on each scale. We therefore apply a similar calculation to compute each component of the velocity field at each scales. The only difference is that the variance is determined directly by the turbulent spectrum (in our case, $\sigma_\mathrm{g}^2(R) \propto R$), rather than via equation~\ref{eqn:sigma.mach.general}. }

\begin{figure}
    \centering
    \includegraphics[width=0.5\linewidth]{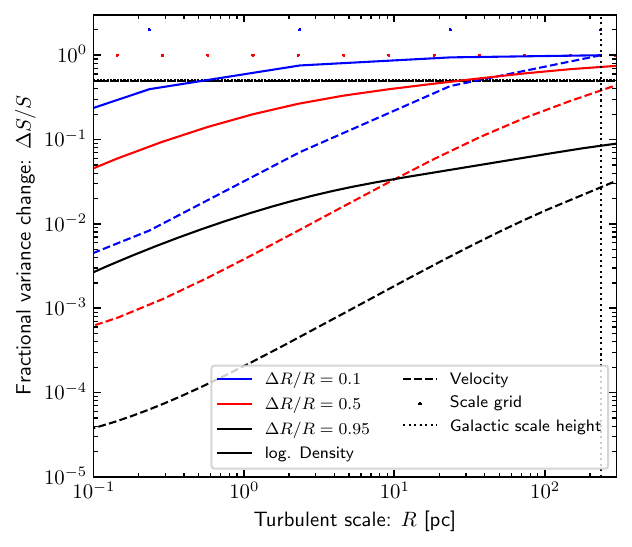}
    \caption{{The fractional change in the variance $\Delta S/S$ of the Gaussian random field at each spatial scale with varying spatial resolution $\Delta R/R$. We mark the scale height $h$ as a vertical dotted line. The location at which each $\Delta S$ is computed are shown as scatter points.}}
    \label{fig:deltaS_S}
\end{figure}

\subsection{Critical density and the mass of collapsing GMCs}
\label{app:crit_density}

In order to compute the mass function of collapsing GMCs, we must determine the critical density at which a GMC undergoes gravitational collapse. Given the dispersion relation for density fluctuations in a disc of finite thickness $h$ \citep[e.g.][]{Begelman09}, it can be shown that unstable collapse occurs if:
\begin{align}
\label{eqn:rhocrit}
\frac{\rho}{\rho_{0}} &\ge \frac{\rho_{\rm c}}{\rho_{0}} = \frac{Q_{0}(h)}{2\,\tilde{\kappa}}\,(1+\tilde{k})
\left[ \frac{\sigma_{\rm g}(k)^{2}}{\sigma_{\rm g}(h^{-1})^{2}}\,\tilde{k}  + \tilde{\kappa}^{2}\,\tilde{k}^{-1}\right] ,
\end{align}where we have defined the dimensionless versions of the epicyclic frequency  $\tilde{\kappa} \equiv \kappa/\Omega =\sqrt{2}$ and wavenumber $\tilde{k}\equiv |k|\,h$. For a disc in vertical equilibrium $\sigma_{\rm g}(h^{-1}) = h\Omega$, and we will fix $\sigma_{\rm g}(h^{-1}) = 6$~km~s$^{-1}$. This is the fiducial case explored by \citet{Hopkins12} and yields a scale height $h=235$~pc similar to the observed thin disc scale height \citep[e.g.][]{Viera23}. Then at any given scale, we can define an analogous critical logarithmic density perturbation:
\be
\delta_{\rm c}(R) = \ln{ \left(\frac{\rho_{\rm c}}{\rho_{0}} \right)} - \left\langle{\ln{\left(\frac{\rho}{\rho_{0}} \right)}} \right\rangle.
\ee The unstable, collapsing star forming region is defined at the largest scale $R_\mathrm{SFR}$ for which $\delta \geq \delta_{\rm{c}}$. Interpolating to the cylindrical geometry if the $R_\mathrm{SFR} \gtrsim h$, the gas mass of the collapsing cloud is defined:
\be
\label{eqn:mass.size}
M(\rho\,|\,R) \equiv 4\,\pi\,\rho(R)\,h^{3}\,
\left[\frac{R^{2}}{2\,h^{2}} + \left(1+\frac{R}{h}\right)\,\exp{\left(-\frac{R}{h}\right)}-1 \right],
\ee where $\rho = \rho_{\rm{c}}$ and $R = R_\mathrm{SFR}$. 

\subsection{Evolution of the turbulent medium}
\label{app:time_evol}

We can compute the temporal evolution of the ISM by assuming the turbulence is globally steady-state, such that the turbulent velocity cascade is maintained outside of collapsing regions. The probability density function for any Gaussian random variable $\delta$ under these conditions obeys a generalized Fokker-Planck equation. We can therefore redefine our trajectory at $t+\Delta t$ by computing a random walk step:
\begin{equation}
\label{eq:Deltadelta_t}
\Delta \delta_{j}(t+\Delta t) =  \Delta \delta_{j}(t)\,\exp{(-\Delta t/\tau_{\rm{t}})}  + \mathcal{R} \, \sqrt{\Delta S\,(1-\exp{(-2\,\Delta t/\tau_{\rm{t}})})},
\end{equation}
where $\mathcal{R}$ is a Gaussian random number with unity variance. The new trajectory is the sum of the components given by equation~\ref{eq:Deltadelta_t}. We define the time-step to be $2$~percent of the shortest turbulent time-scale in our grid. Equation~\ref{eq:Deltadelta_t} applies both to the logarithmic density perturbations and to each component of the interstellar gas velocity.

\section{Cloud collapse and star formation rate}
\label{app:collapse}
\restartappendixnumbering
In the early stage after a star forms, it occupies the local star forming region that is undergoing gravitational collapse. Since we are interested in the local density and velocity evolution throughout a star's lifetime, we require a simple model to follow the collapse of the GMC and formation of stars. Once the GMC becomes gravitationally unstable, it collapses on a free-fall time-scale. It does not do so monolithically, but because smaller sub-regions of the cloud may be higher density (and therefore have shorter free-fall time-scales), they fragment first and continue to accrete from their parent structures \citep{Hoyle53}. Simultaneously, the cloud is exposed to heating by feedback processes from the stars as they form. A wide range of analytic and semi-analytic approaches have been developed to describe this process \citep[see][and references therein]{Vazquez-Semadeni19}. We are helped in this context in our interest in low mass star forming regions that are locally well sampled. Since we do not expect many massive stars or supernovae among these local regions, we will hereafter ignore feedback processes and exclusively consider gravitational collapse. 

We adopt the approach of \citet{Girichidis14} in modelling the rate of star formation during the collapse of the GMC. The approach of those authors is attractive both for its simplicity and the self-consistent treatment of the gas density distribution as it evolves under gravity. We briefly review the approach here, referring the reader to the original work for details. We consider a spherical region undergoing free-fall collapse from rest, which may be contained within a larger gravitationally unstable cloud. The outer radius radius $r$, obeys the equation:
\begin{equation}
  \label{eq:free-fall-DGL}
  \frac{\dif^2 r}{\dif t^2} = -\,\frac{GM}{r^2} = -\,\frac{4\pi G \rho_1 r_1^3}{3r^2}\,.
\end{equation}
Here, $M = \rho V = (4 \pi/3)  \rho r^3$ is the fixed mass of the region with volume $V$, initial density $ \rho_{1} = \rho(0)$ and radius $r_{1} = r(0)$. We have dropped the flattened geometry of equation~\ref{eqn:mass.size} for simplicity, but also because we are considering low mass (small) regions that have $r_{1}<h$. We define the dimensionless time:
\begin{equation}
    \theta = t/\tau_\mathrm{ff},
\end{equation}where
\begin{equation}
  \tau_\mathrm{ff} = \sqrt{\frac{3\pi}{32 G \rho_1}\,} = \left(\gamma \rho_1 \right)^{-1/2},
\end{equation}is the free-fall time, and the second equality defines $\gamma$ for convenience. While there is no exact solution to equation~\ref{eq:free-fall-DGL}, \citet{Girichidis14} point out that 
\begin{equation}
  \label{eq:approx-radius-solution}
  r (\theta) = r_{1}\left({1-\theta^2}\right)^{p/3},
\end{equation}with $p=2$, is a good approximation. If the mass is constant, then the density evolves as:
\begin{equation}
  \label{eq:approx-density-solution}
  \rho(\theta)=\rho_{1}\rkl{1-\theta^2}^{-p}.
\end{equation}

To determine the rate at which mass is accreted onto forming stars, we compute the rate at which mass evolves above some threshold density for accretion $\rho_\mathrm{a}$. We will choose $\rho_\mathrm{a}$ to yield a free-fall time $0.1$~Myr, although this choice does not strongly influence the star formation rate as long as this time-scale is much shorter than the global free-fall time-scale. The problem is now reduced to finding the initial density $\rho_*$ that evolves to $\rho_\mathrm{a}$ at time $t$. If $p\!=\!2$, then inverting equation~\ref{eq:approx-density-solution}:
\begin{equation}
\rho_*(\rho_{\rm{a}},t) = \frac{2\gamma \rho_{\rm{a}} t^2 + 1}{2\gamma^2 \rho_{\rm a} t^4} + \sqrt{\rkl{\frac{2\gamma \rho_{\rm a} t^2 + 1}{2\gamma^2 \rho_{\rm{a}} t^4}}^2 - \frac{1}{\gamma^2 t^4}}.
\end{equation}

Now we just need to calculate the fraction of the total cloud mass that has $\rho>\rho_*$. Given that we have an initial probability density function in terms of volume (as in equation~\ref{eqn:lognormal}):
\begin{equation}
  \pdfv = \frac{1}{V_\mathrm{tot}} \frac{\dif V}{\dif\rho},
\end{equation}we can convert this into the mass-weighted version:
\begin{equation}
  \pdfm = \frac{1}{M_\mathrm{tot}} \frac{\dif M}{\dif\rho},
\end{equation}via the equation:
\begin{equation}
  \label{eq:conversion-mass-to-volume-weighting}
  \frac{\dif V}{\dif\rho} = \frac{\dif V}{\dif M}\frac{\dif M}{\dif\rho} = \frac{1}{\rho}\frac{\dif M}{\dif\rho},
\end{equation}for some total volume $V_\mathrm{tot}$ or mass $M_\mathrm{tot}$. Here we define $M_\mathrm{tot} = M(\rho_\mathrm{c}|R_\mathrm{SFR})$ from equation~\ref{eqn:mass.size}. Now the total mass in stars is just:
\begin{equation}
    M_* = M_\mathrm{tot} \int_{\rho_*}^\infty P_M(\rho) \,\mathrm{d}\rho = 
    \frac{M_\mathrm{tot}}{2} 
    \ekl{ 1+ \operatorname{erf}\rkl{\frac{\ln 
    \frac{\rho_\mathrm{c}}{\rho_*} + \sigma_k^2 }
    {\sqrt{2}\sigma_k}} },
\end{equation}where the last equality assumes an initially lognormal probability density distribution in volume. As in Section~\ref{sec:local_sfrs}, the critical density of the region is $\rho_\mathrm{c}(R_\mathrm{SFR})$ and $\sigma_k$ is the dispersion given $k \sim 1/R_\mathrm{SFR}$, for $R_\mathrm{SFR}$ the largest scale of the collapsing star forming region. We stop star formation and assume the cloud is immediately dispersed once $M_* = \epsilon M$, where we adopt $\epsilon=0.5$ consistent with calculations for core formation \citep{Matzner00}. 

We show the evolution of some example star forming regions according to this simple model in Figure~\ref{fig:density_SFR}. Over the typical range of the masses $M_\mathrm{SFR}$ of the star forming regions that we consider ($10-500 \, M_\odot$ in our fiducial model), the mean density of the regions when the majority of stars form decreases by an order of magnitude. However, the period of time for which stars typically inhabit the collapsing cloud increases from a few $10^5$~years for $M_\mathrm{SFR} \sim 10\, M_\odot$ to $\sim 1$~Myr at $M_\mathrm{SFR} \sim 500\, M_\odot$. Given that the cumulative amount of mass accreted onto stars is proportional to density and time, we then expect a fairly weak dependence on the \textit{total} mass accumulated by discs during the bound, star forming phase.

\begin{figure}
    \centering
    \includegraphics[width=0.5\linewidth]{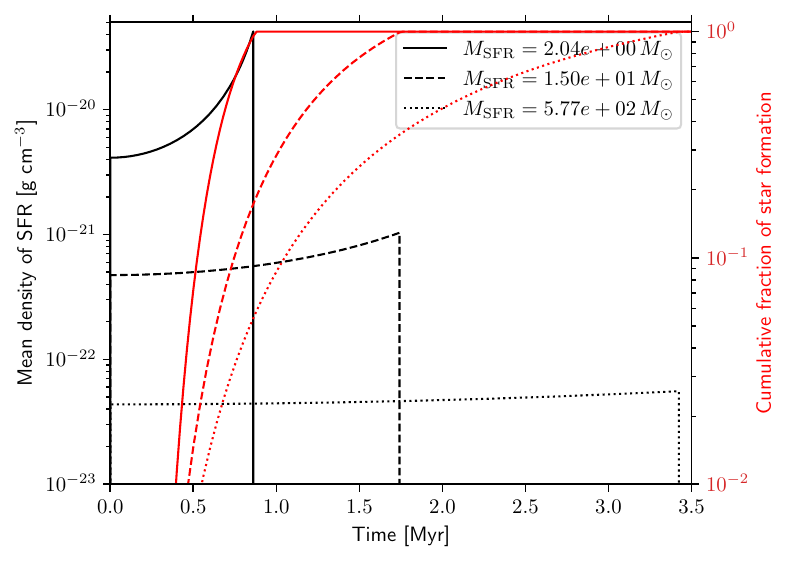}
    \caption{Evolution of the mean density (black lines, LHS axis) and fraction of star formation in three star forming regions (red lines, RHS axis) of different mass in our model. We show the results for three different initial masses of the star forming regions, $M_\mathrm{SFR}$. }
    \label{fig:density_SFR}
\end{figure}

%In order to maintain the simplicity of this approach, we have had to assume that none of the mass of the cloud is lost during the star formation phase, all mass going into stars that also collapse. Assuming the global collapse remains similar, deferring the mass removal due to outflows only alters the density evolution of the region by a factor of order unity. However, we have also missed numerous physical mechanisms that contribute to the molecular cloud growth, collapse, star formation and feedback-driven dispersal \citep[e.g.][]{Vazquez-Semadeni19}. Nonetheless, we expect our minimalist approach to capture the essential physics for the problem at hand; in particular, the time-scale for star formation rate and the local gas density evolution for low mass star forming regions. For example, our choice to ignore cloud growth is mitigated in that we start the collapse \textit{after} the cloud has already grown to become unstable. As another example, while we ignore feedback from stellar winds and supernovae, we consider only low mass star forming regions that should have little feedback and a high star formation efficiency \citep[see figure 7 in][]{Vazquez-Semadeni19}. In the context of this work, the value of simplicity outweighs the benefits of including more processes at the cost of free-parameters or model choices.   

\section{Sensitivity to disc dispersal criterion}
\label{app:discdisp_dprops}
\restartappendixnumbering
\begin{figure}
    \centering
    \includegraphics[width=0.8\linewidth]{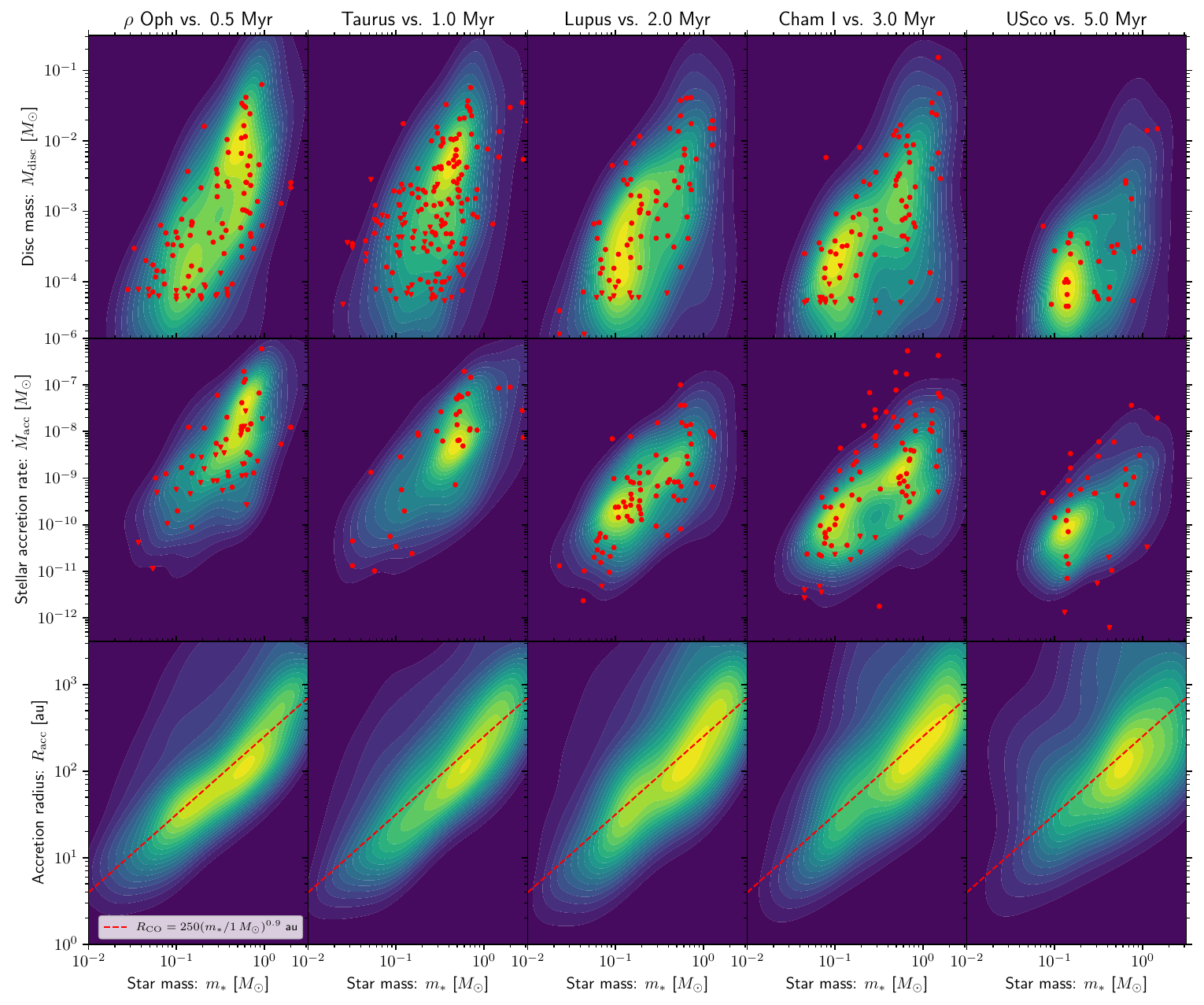}
    \caption{As in Figure~\ref{fig:discprops}, but with a reduced dispersal disc mass threshold, $M_\mathrm{disc} < 10^{-5} \, M_\odot$}
    \label{fig:discprops_ll-5}
\end{figure}

We have used a fiducial criterion for disc dispersal (when $M_\mathrm{disc}<3\times 10^{-5} \,M_\odot$). While few disc-hosting stars have been found with nominal mass subtantially lower than this, the choice is to some degree arbitrary. Here we consider how reducing this threshold by a moderate factor of three alters the distribution of disc properties and replenishment fractions. 

We show a comparison to Figure~\ref{fig:discprops} with $M_\mathrm{disc}>10^{-5} \, M_\odot$ in Figure~\ref{fig:discprops_ll-5}. There are some slight differences in the distributions of disc properties, but they remain qualitatively similar to the observed distributions. For example, the typical accretion radius is slightly smaller in the low mass threshold case. This is because the accretion radius for the very low mass discs may be set by a period of BHL accretion with comparatively low BHL accretion rates (corresponding to a smaller BHL radius). In Figure~\ref{fig:hlrf_ll-5}, we also compare the replenishment fractions to those in Figure~\ref{fig:hlrf}. We find a small increase in the replenishment fraction for discs, particularly at late times. This is because discs with a higher fraction of replenished material are often lower mass. In summary, we find only minor differences in our model outcomes if we reduce the disc dispersal criterion by a factor few.

\begin{figure}
    \centering
    \includegraphics[width=0.8\linewidth]{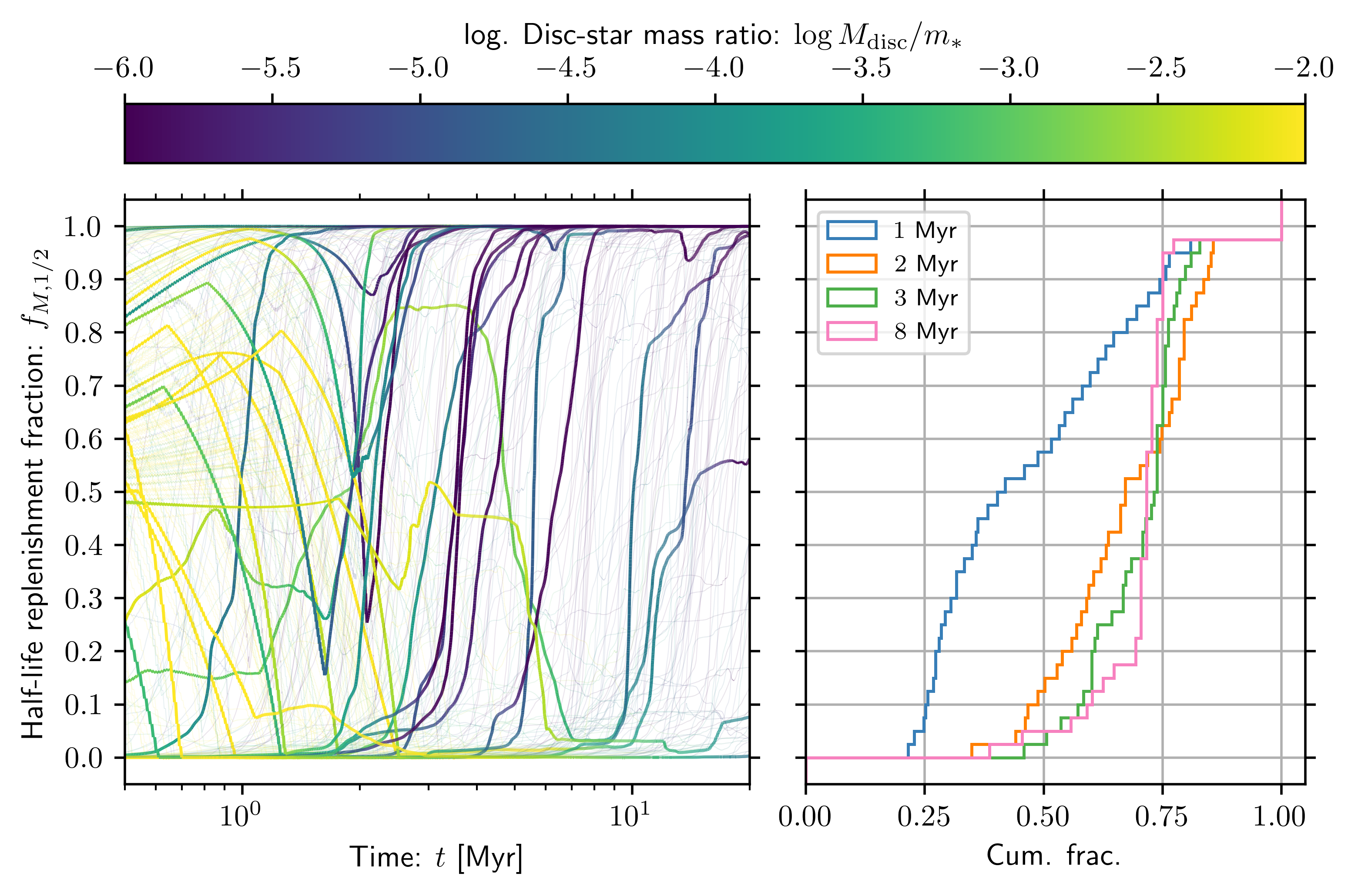}
    \caption{As in Figure~\ref{fig:hlrf}, but with a reduced dispersal disc mass threshold, $M_\mathrm{disc} < 10^{-5} \, M_\odot$. }
    \label{fig:hlrf_ll-5}
\end{figure}

\section{Sensitivity to the initial disc mass:} 

\restartappendixnumbering
\label{app:initial_dmass}

We consider how our conclusions on disc replenishment as a function of time change when we do not assume a negligible initial disc mass, as we assumed in our fiducial model. We therefore repeat our calculations with disc masses initially similar to the observed distribution in young star forming regions. In practice, we adopt a log-normal distribution for the disc mass with $1$~dex of scatter around the relationship 
\begin{equation}
    M_\mathrm{disc}(t=0) = 0.01 \left( \frac{m_*}{1\, M_\odot}\right)^2 \, M_\odot.
\end{equation}We show the resultant disc properties in ~\ref{fig:discprops_minit_0.01}, which again are in good agreement with the observed disc properties. Our adopted initial disc masses are approximately the maximum we could adopt while still remaining consistent with the observed distribution of disc masses. The outcome for the half-life disc replenishment factor is shown in Figure~\ref{fig:hlrf_minit_0.01}. Although in this case $f_\mathrm{M,1/2}$ is somewhat reduced early on in the disc evolution, overall differences in the distribution over time is minor. The fraction of discs with at least some recently accreted material remains similar, while we find that $f_\mathrm{M,1/2} \sim 0.2{-}0.45$ throughout the disc lifetime. Therefore, even under the assumption that the primordial disc is coincidentally similar in mass to the subsequently accreted one, and ignoring all internal depletion processes, we still expect a substantial fraction of discs to be composed of recently captured disc material. Our conclusion that the present day disc properties are not a good probe of internal disc physics therefore appears robust. 

\begin{figure}
    \centering
    \includegraphics[width=\textwidth]{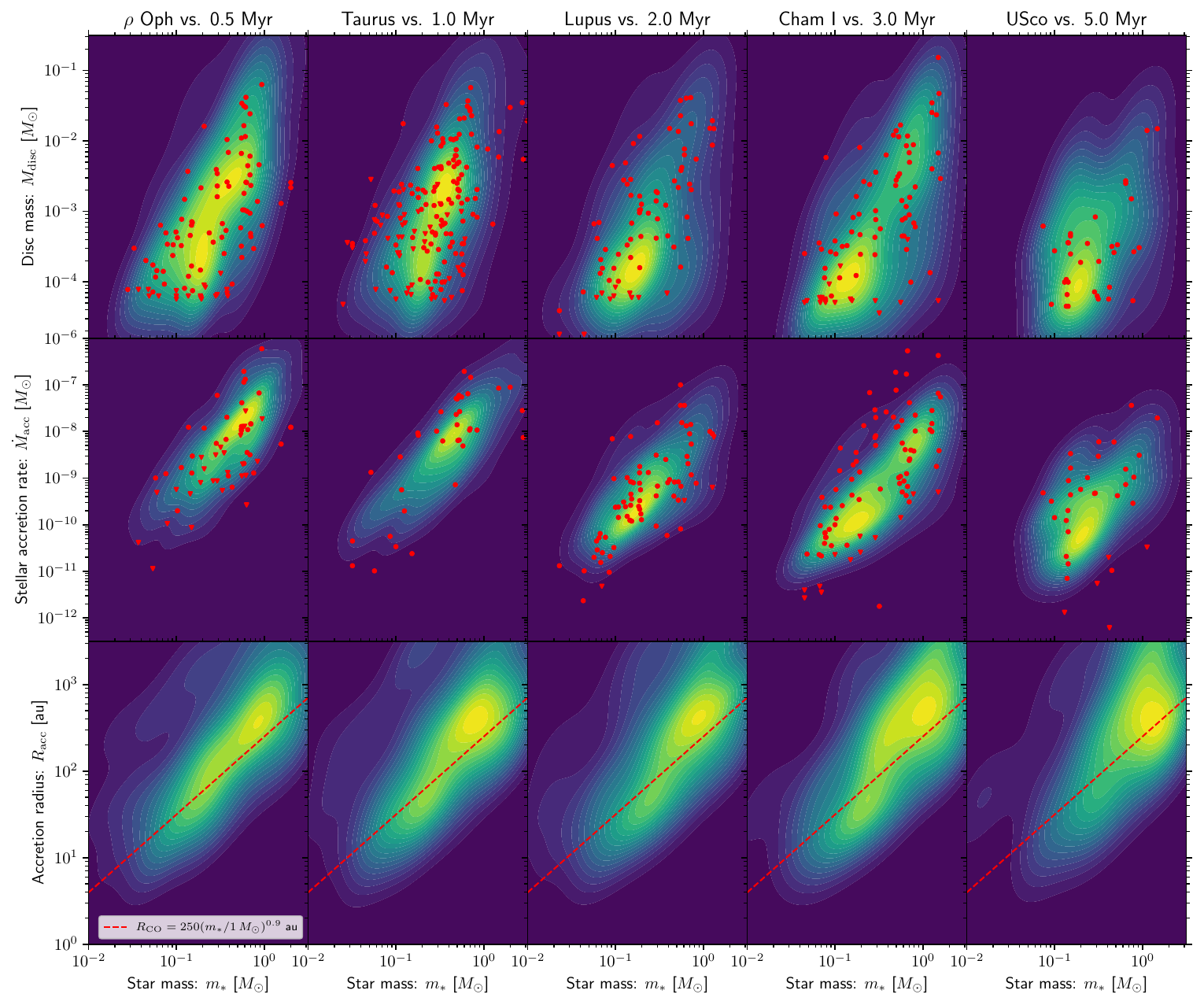}
    \caption{\label{fig:discprops_minit_0.01}As in Figure~\ref{fig:discprops}, but for an initial disc mass $M_\mathrm{disc}(t=0) = 0.01(m_*/1\, M_\odot)^2 \, M_\odot$ with $1$~dex of scatter.}
\end{figure}

\begin{figure}
    \centering
    \includegraphics[width=\columnwidth]{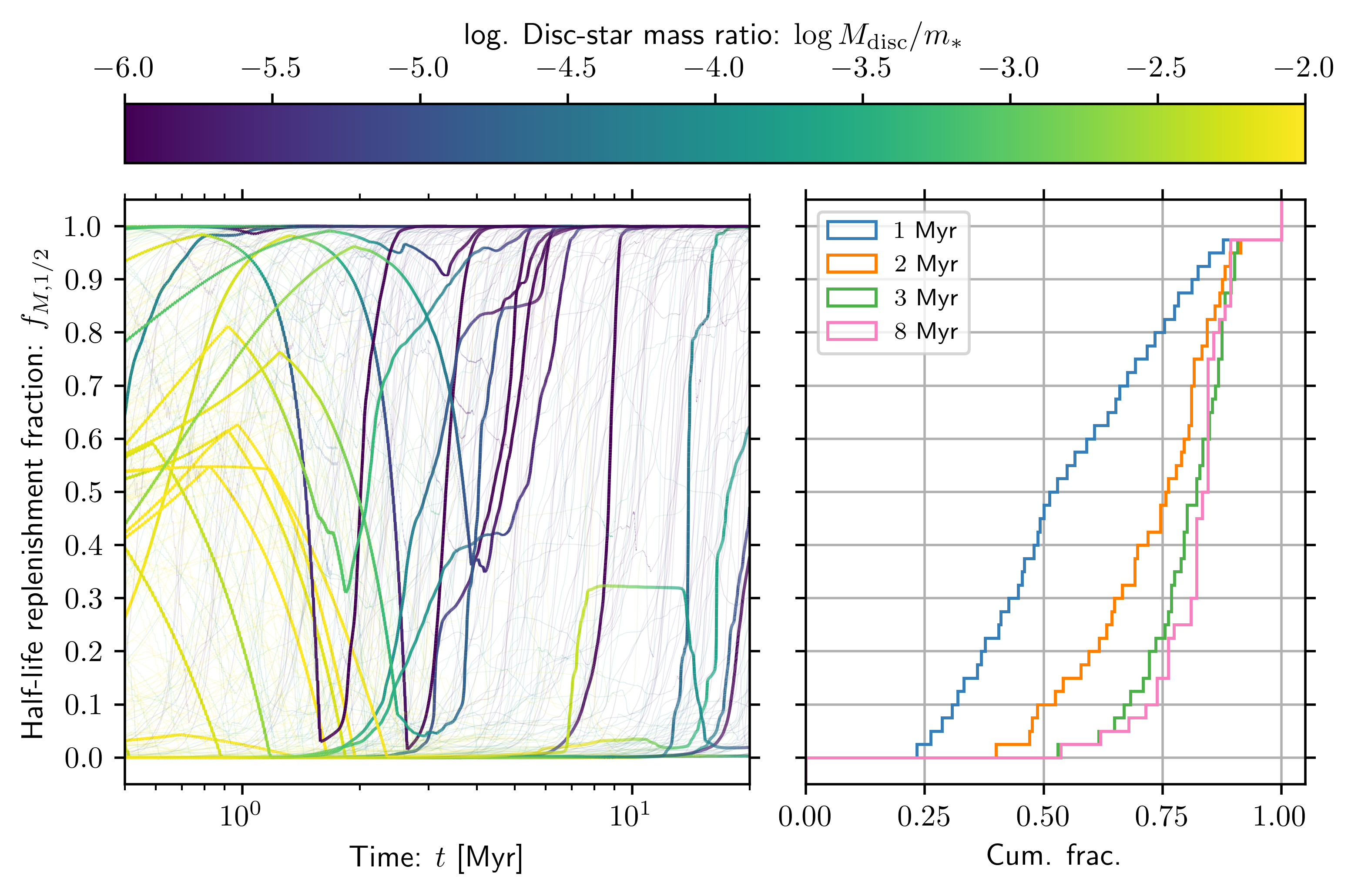}
    \caption{\label{fig:hlrf_minit_0.01}\small{As in Figure~\ref{fig:hlrf}, but for an initial disc mass $M_\mathrm{disc}(t=0) = 0.01(m_*/1\, M_\odot)^2 \, M_\odot$ with $1$~dex of scatter.}}
\end{figure}

\section{Stellar feedback and BHL accretion}
\label{app:feedback_BHL}

\restartappendixnumbering

\subsection{Feedback overview}

Several authors have explored how feedback during massive star formation can alter the temperature in accretion flows, and therefore reduce the efficiency of mass accretion from the ISM \citep{Wolfire86, Edgar03, EdgarClarke04}. For massive stars, when dust is sublimates and high temperatures are reached, the accretion flow can be strongly perturbed or suppressed. In our case, we focus on much lower stellar masses and weaker irradiation, such that we do not expect dust sublimation in the flow. However, we are also interested in much lower accretion rates and lower local ISM density than expected during the formation of massive stars. In some cases, we then expect feedback to limit accretion. Understanding the role of feedback requires a devoted study, but here we present a simple estimate for the degree to which feedback limits BHL accretion flow. First (Section~\ref{sec:rtherm}), we consider passive heating of the accretion flow, which we show is never a stringent limit on the BHL accretion rate. More importantly, in Section~\ref{sec:rwind}, we consider suppression of the accretion flow through photoevaporative winds, which are launched by high energy photons. Both of these mechanisms depend on the evolution of the feedback from the central star, so we first briefly outline our stellar model that use to compute the evolution of the stellar properties.

\subsection{Stellar evolution models}

To compute the feedback efficacy, we require $L_*$ given a stellar mass $m_*$ of a certain age. For this, we use the pre-computed \textit{MESA Isochrones and Stellar Tracks} \citep[MIST\footnote{\url{https://waps.cfa.harvard.edu/MIST/}} --][]{Paxton11, Dotter16, Choi16}. When we consider high energy photons, we also require a stellar spectrum. We therefore couple the MIST isochrones with the stellar atmosphere models of \citet{Cas04}. However, these models do not account for the accretion luminosity that can dominate energy output at short wavelengths for low mass stars. As a simple estimate of this accretion luminosity, we will assume:
    \begin{equation}
        L_\mathrm{acc} = \frac{G m_* \dot{M}_\mathrm{acc}}{R_*},
    \end{equation}where $\dot{M}_\mathrm{acc}$ is the accretion rate onto the star (not to be confused with $\dot{M}_\mathrm{BHL}$), and $R_*$ is the stellar radius. We assume that the accretion flow emits as a blackbody with temperature:
    \begin{equation}
        T_\mathrm{acc} = \left( \frac{L_\mathrm{acc}}{4\pi R_*^2 \sigma_\mathrm{SB}}\right)^{1/4}.
    \end{equation}When we come to consider the short wavelength ($\lambda< 2070$~\r{A}) photons that drive stellar winds, we will integrate over the stellar spectrum:
    \begin{equation}
        L_\mathrm{*,<2070} = \int_0^{2070} L_{*,\lambda} \,\mathrm{d}\lambda,
    \end{equation}where $ L_{*,\lambda}$ is the wavelength-specific stellar luminosity.

\subsection{Thermal radius}
\label{sec:rtherm}
 We now consider whether, given the thermal structure of the flow imposed by stellar irradiation, whether the temperature is ever sufficient to raise the sound speed $c_\mathrm{s}>c_\mathrm{s,crit} \sim v_\mathrm{esc}$, where $v_\mathrm{esc}$ is the escape velocity. In such a scenario, the gas at the stagnation point would not be bound to the star, and would therefore escape rather than undergo capture. To estimate the degree to which passive heating by the star can unbind the accretion flow, we will assume that the flow is instantaneously at rest at the accretion radius $R_\mathrm{acc}$. The problem is then very similar to a passively irradiated protoplanetary disc \citep{Chiang97}. The surface of the column is heated at a rate $Q_\mathrm{irr}$, while the disc cools at a rate $Q_\mathrm{cool}$ at the surface of the accretion column. The different geometry of the accretion flow (for example, a cone that widens far from the star) compared to the disc just means that both the heating and cooling areas at a given radius are increased by the same factor. Then the centre of the column has temperature $T_\mathrm{col}$ such that:
    \begin{equation}
        \sigma_{\mathrm{SB}} T_\mathrm{col}^4 = \frac{\phi L_*}{8\pi R_\mathrm{acc}^2 }.
    \end{equation}Here $L_*$ is the bolometric luminosity, and $\phi$ in the context of a protoplanetary disc is the flaring angle -- i.e. the angle at which the stellar radiation meets the surface structure. The largest value this can be is $\phi=\pi/2$, which would correspond to a column where the ISM interior to $R_\mathrm{acc}$ is optically thin. Although in this case our assumptions on the column geometry are broken, we adopt this as the most restrictive possible case. 
    
    It is clear from the above discussion that $c_\mathrm{s}/c_\mathrm{s, crit} \propto R_\mathrm{acc}^{1/4}$, so that there always formally exists some radius for which $c_\mathrm{s}>c_\mathrm{s, crit}$ (although in practice this radius may be sufficiently far away such that the central star does not set the ISM temperature). Therefore the accretion radius $R_\mathrm{acc}$ can always therefore be limited by $R_\mathrm{therm}$. The outcome of these calculations are shown in Figure~\ref{fig:Rtherm}. We find that $R_\mathrm{therm}$ is typically much smaller for low mass stars, but that even for $m_* = 0.1 \, M_\odot$ we have $R_\mathrm{therm} \sim 5\times 10^4$~au$~
    \sim R_\mathrm{T}$, the tidal radius. This means that $R_\mathrm{therm} \gtrsim R_\mathrm{T}$ in most scenarios, and we do not expect the passive stellar irradiation to limit BHL accretion. 

\begin{figure}
    \centering
    \includegraphics[width=0.6\textwidth]{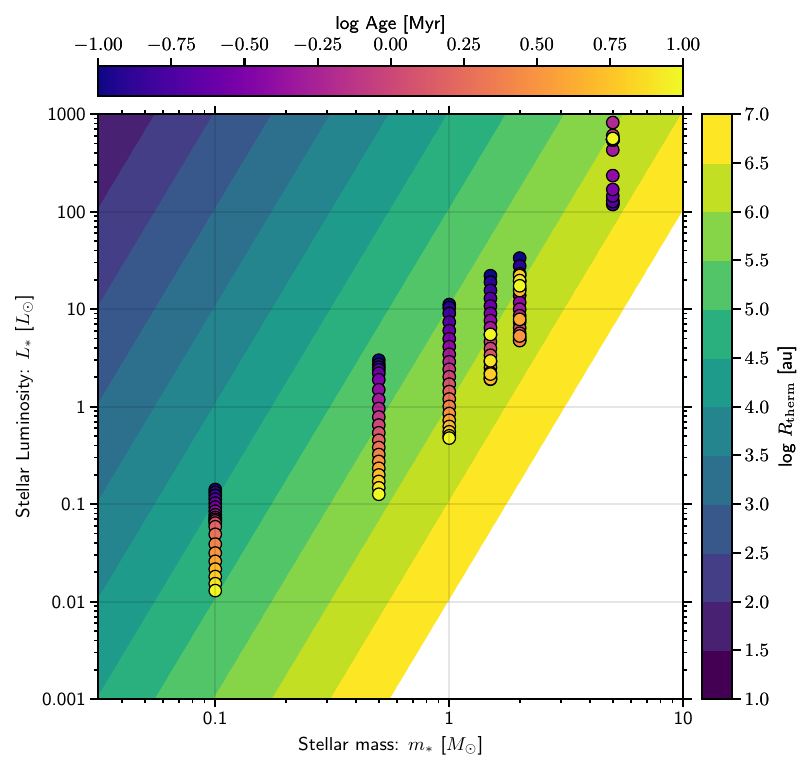}
    \caption{The thermal radius $R_\mathrm{therm}$, defined to be the radius at which the bolometric stellar luminosity heats the gas sufficiently be able to escape the gravity of the central star, as a function of stellar mass and luminosity. We also show the time-evolution for stars of different initial stellar mass in the $m_*-L_*$ plane. These points are coloured by the age of the star.}
    \label{fig:Rtherm}
\end{figure}

\subsection{Internally driven wind}

\subsubsection{Wind radius}
\label{sec:rwind}
A second feedback effect is the photoevaporative wind that can be driven by ionising or photodissociating photons. These photons can heat the outer layers of the accretion flow, and drive mass-loss in a thermal wind that may be comparable to the BHL accretion rate. In principle, computing the wind mass-loss rate $\dot{M}_\mathrm{wind}$ requires considering the full microphysics within the photodissociation region \citep[e.g.][]{Haworth18, Haworth23}. Perhaps even more importantly, as discussed by \citet{EdgarClarke04}, the highly tangential direction of the streamlines as they approach the star makes the flow vulnerable to the momentum injected from within. A proper prescription of the role of feedback on the BHL accretion flow for low mass ($\lesssim 3\, M_\odot$) stars will require a devoted study. We nonetheless present a crude approximation as follows. 

We base our estimate on injecting a fixed fraction $\epsilon_\mathrm{wind}$ of energy from high energy photons (wavelength $\lambda < 2070 $~\r{A}) into unbinding mass at a velocity $v_\mathrm{launch}$ from the BHL radius. As discussed above, this estimate is crude in that it ignores the geometry of the flow. For the purpose of this work, it should be considered an order of magnitude estimate of the conditions under which winds become important. We will assume $v_\mathrm{launch} \sim v_\mathrm{esc} = \sqrt{2G m_*/R_\mathrm{acc}}$. In this case:
    \begin{equation}
        \frac 1 2 \dot{M}_\mathrm{wind} v_\mathrm{launch}^2 =\epsilon_\mathrm{wind} L_\mathrm{*, <2070}  \iff \dot{M}_\mathrm{wind} =\epsilon_\mathrm{wind}\frac{L_\mathrm{*, <2070}   R_\mathrm{acc}}{Gm_*}.
    \end{equation}BHL accretion can only proceed if $\dot{M}_\mathrm{BHL} > \dot{M}_\mathrm{wind}$, or:
    \begin{equation}
          R_\mathrm{acc}  \gtrsim \left({2 G m_*}\right)^{-3}\left(\frac{\epsilon_\mathrm{wind} L_\mathrm{*, <2070}}{\pi   \rho}\right)^2 = R_\mathrm{wind}.
    \end{equation}The factor $\epsilon_\mathrm{wind}$ folds in the fractional solid angle covered by the accretion flow and the complex wind physics. Hereafter we will assume $\epsilon_\mathrm{wind} = 0.1$. We have also assumed that the wind is launched at the minimum possible velocity $v_\mathrm{esc}$, so that $10$~percent of the available energy from high energy photons goes into launching the maximal possible mass in the wind. We expect this to be a conservative assumption from the perspective of maximising replenishment. In practice, many photons will escape along low density lines-of-sight (reducing $\epsilon_\mathrm{wind}$), and higher energy photons may also launch more rapid winds with velocity $\sim 10$~km~s$^{-1}$. In making this choice, we therefore explore the opposite extreme compared to the `no wind' scenario of our fiducial model.

\begin{figure}
    \centering
    \includegraphics[width=0.6\textwidth]{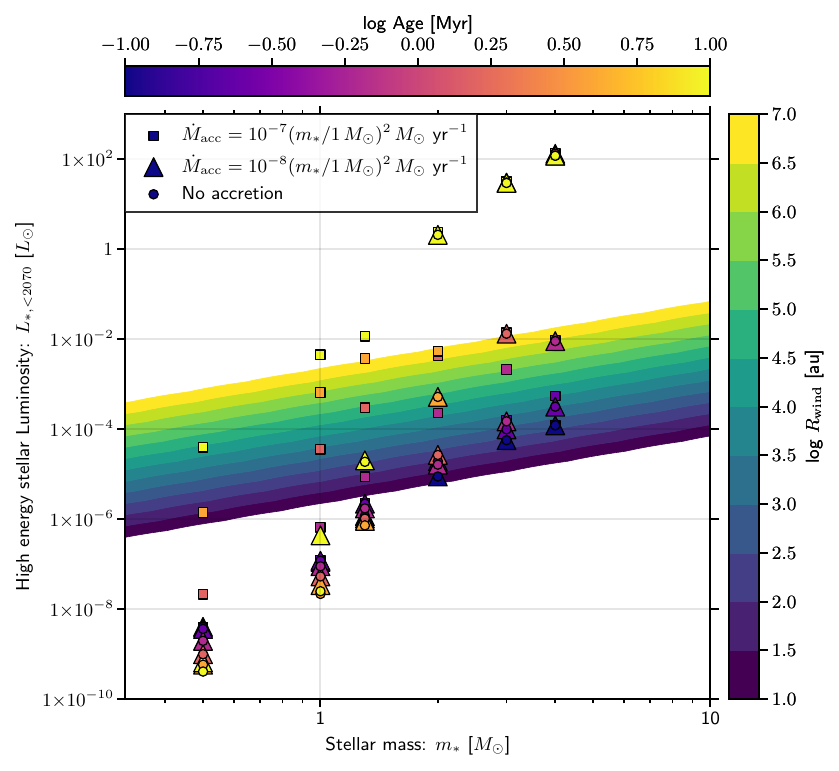}
    \caption{The radius $R_\mathrm{wind}$ inside which the accretion flow may be suppressed by the wind driven by high energy photons, as a function of stellar mass and stellar luminosity at wavelengths $\lambda< 2070$~\r{A}. Points represent the outcome of stellar models with no stellar accretion (circles), accretion rates following $\dot{M}_\mathrm{acc}= 10^{-8}(m_*/1\,M_\odot)^2 \, M_\odot$~yr$^{-1} $ (triangles) and  $\dot{M}_\mathrm{acc}= 10^{-7}(m_*/1\,M_\odot)^2 \, M_\odot$~yr$^{-1} $ (squares). These points are coloured by the age of the star.}
    \label{fig:Rwind}
\end{figure}

\subsubsection{BHL accretion rates with wind suppression}

In order to estimate the role of internally driven photoevaporative winds on limiting the BHL accretion flow, we make the assumption that no accretion cannot proceed if $R_\mathrm{acc}<R_\mathrm{wind}$. We assume $\dot{M}_\mathrm{BHL} = 0$ if this condition is met. In practice, for efficiency we pre-compute a table of $R_\mathrm{wind}$ in stellar age-mass space, and then interpolate and re-scale for the instantaneous ambient density. We show examples of the evolution of the BHL accretion rate resulting from this prescription in Figure~\ref{fig:mdotevol_wind}, which can be compared to Figure~\ref{fig:tevol}. The key difference is that when the ambient gas density is low, instead of experiencing a low $\dot{M}_\mathrm{BHL}$, the star accretes no mass at all. We highlight that physically this would not necessarily be the case. Material outside of the BHL radius may also be hot and/or dense, such that pressure gradients would not necessarily drive gas away from the star. In this case, the star may continue to accrete, either by BHL accretion or, if the BHL radius is inside the disc radius, then by the sweep-up of gas (see Section~\ref{sec:sweep-up}). However, quantifying this process in a star forming region that may be partially ionised on large scales is a complex problem. We neglect this consideration here for the sake of simplicity and in the spirit of a conservative estimate for the BHL accretion rate. 

For the example cases shown in Figure~\ref{fig:mdotevol_wind}, we get a sense that photoevaporation particularly influences the accretion rates for more massive stars (as expected). For our overall sample (which is not drawn from a physical initial mass function or observationally representative equivalent), the result is that the median accretion rate quickly becomes negligible on a time-scale of $\sim 1$~Myr. However, the mean remains relatively unaffected, since the periods of high accretion are not suppressed by the wind. As before, we require disc evolution calculations to understand the degree to which BHL shapes the disc properties over its life-time. 

\begin{figure}
    \centering
    \includegraphics[width=0.5\linewidth]{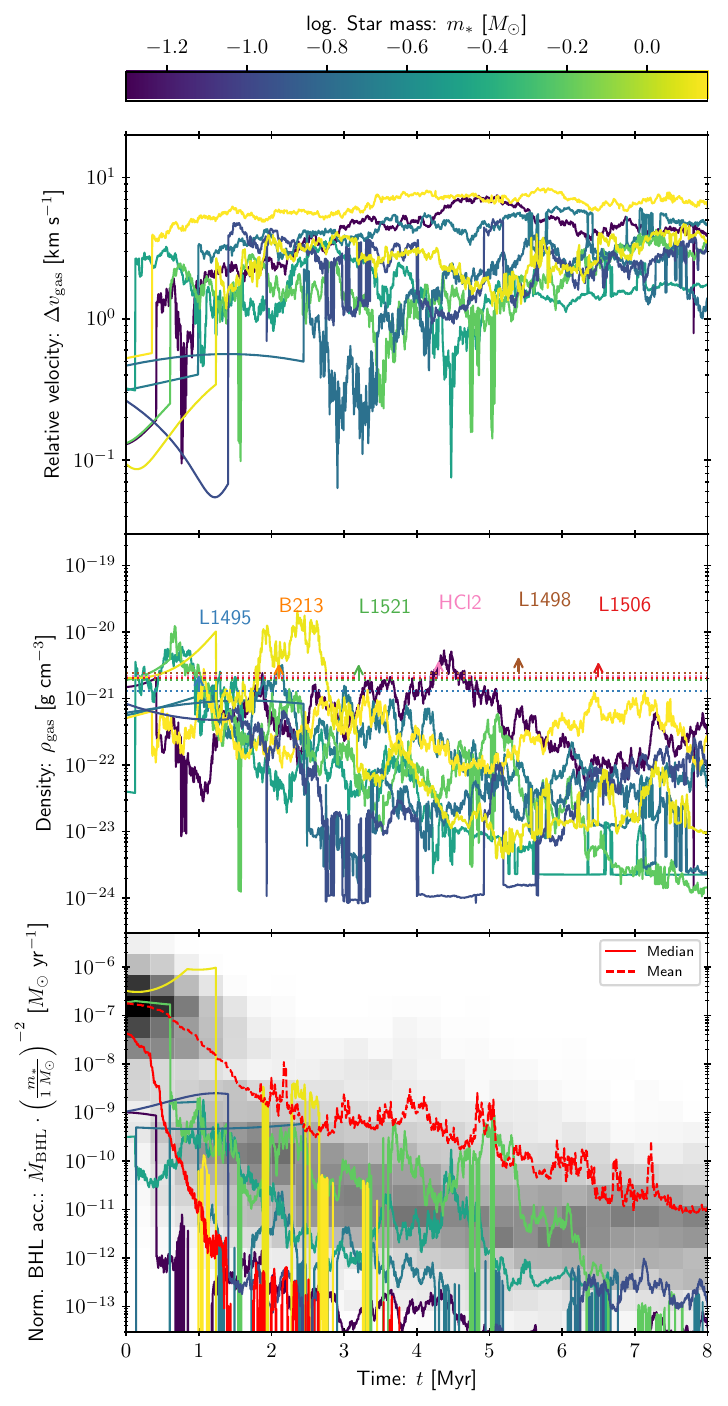}
    \caption{As in Figure~\ref{fig:tevol}, but including a prescription for the suppression of the BHL accretion by internally-driven photoevaporative winds.}
    \label{fig:mdotevol_wind}
\end{figure}

\subsubsection{Disc evolution including winds}

\begin{figure}
    \centering
    \includegraphics[width=0.7
    \linewidth]{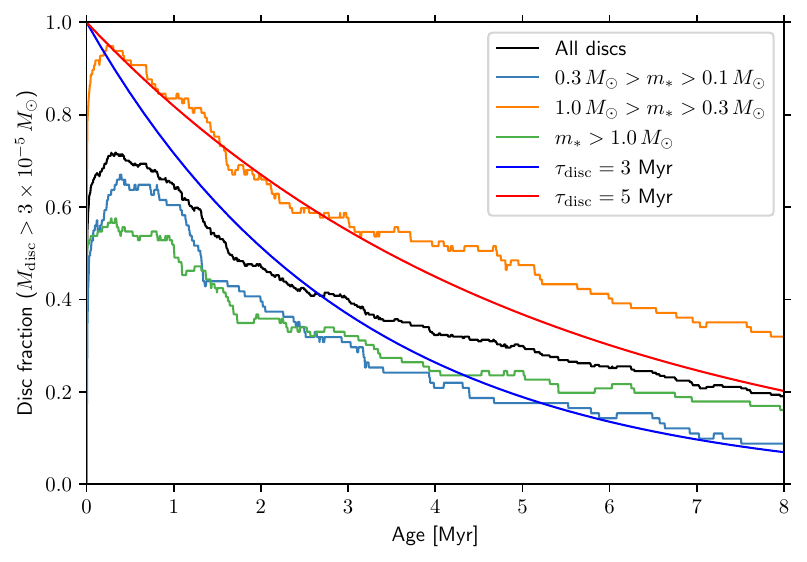}
    \caption{As in Figure~\ref{fig:disc_fraction}, but including a prescription for the suppression of the BHL accretion by internally-driven photoevaporative winds. We have also separated the disc survival fraction up into different mass bins, shown by different line colours. Black represents all stars in our sample with $m_*>0.1\, M_\odot$.}
    \label{fig:discfrac_wind}
\end{figure}

We calculate the disc evolution following the prescription described in Section~\ref{sec:disc_evol}. For numerical reasons, solving the initial value problem is achieved by weighting the BHL accretion rate by a factor $\exp(-R_\mathrm{wind}^2/R_\mathrm{acc}^2)$, to avoid discontinuities in the second derivative. We show the disc fraction resulting from our calculation, analogous to that in Figure~\ref{fig:disc_fraction}, is shown in Figure~\ref{fig:discfrac_wind}. Predictably, the overall disc fraction is somewhat reduced by a moderate factor. However, the influence of winds depends strongly on stellar mass. Dividing the disc lifetimes into difference mass bins, we find that discs with host stars of mass $0.3 \, M_\odot < m_* < 1 \, M_\odot$, which make up a large fraction of the observed samples, show a similar disc fraction evolution to the full sample shown in Figure~\ref{fig:disc_fraction}. Discs around higher mass stars are depleted more rapidly, since BHL accretion is suppressed at a lower ambient density. Stars with $m_*<0.3\, M_\odot$ are again depleted more rapidly, but this is due to our fixed mass threshold for a disc to be considered dispersed. Low mass stars host lower mass discs, with masses that are more frequently below this threshold. 

\begin{figure}
    \centering
    \includegraphics[width=0.9\linewidth]{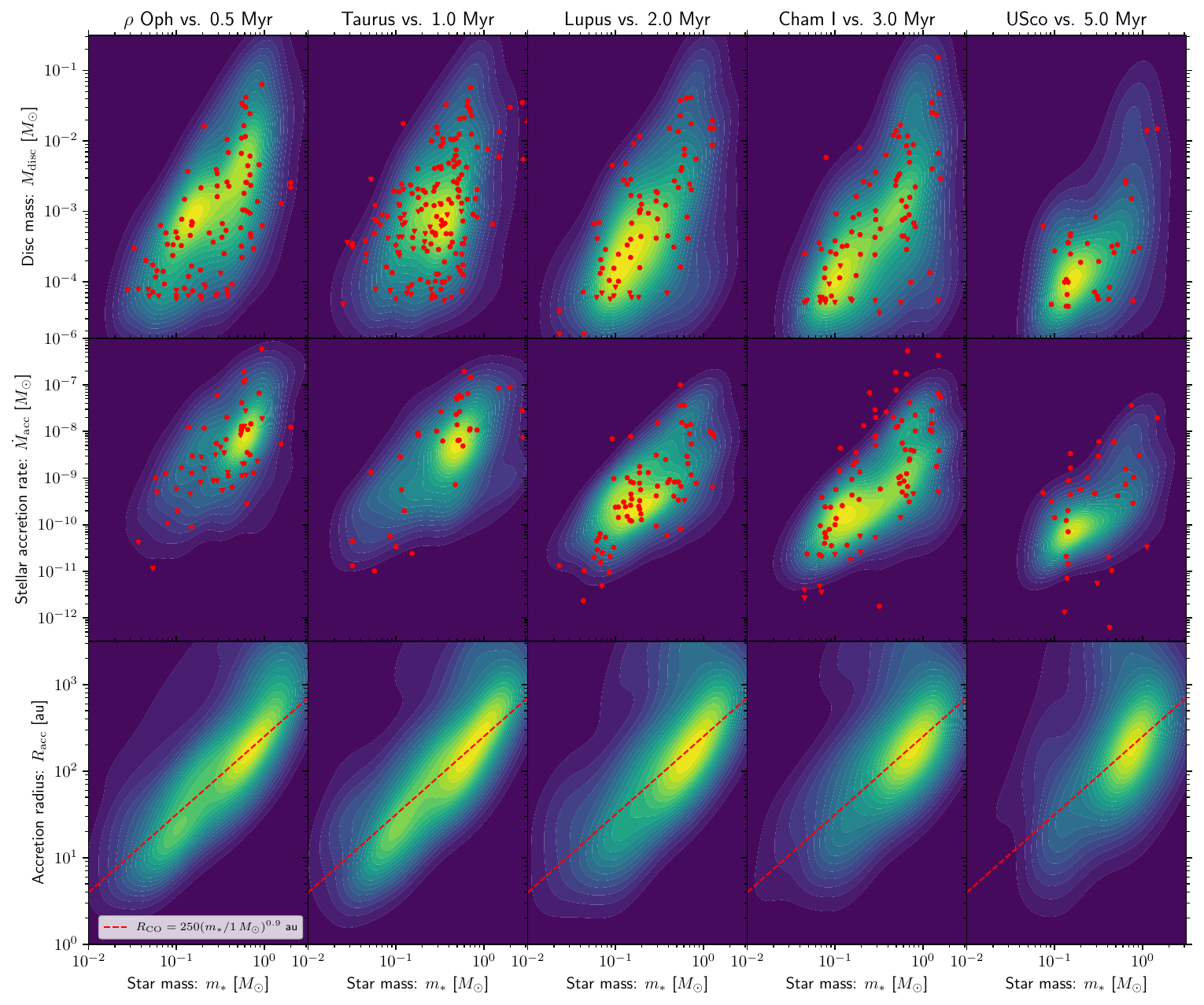}
    \caption{As in Figure~\ref{fig:discprops}, but including a prescription for the suppression of the BHL accretion by internally-driven photoevaporative winds.}
    \label{fig:discprops_wind}
\end{figure}

\begin{figure}
    \centering
    \includegraphics[width=0.9\linewidth]{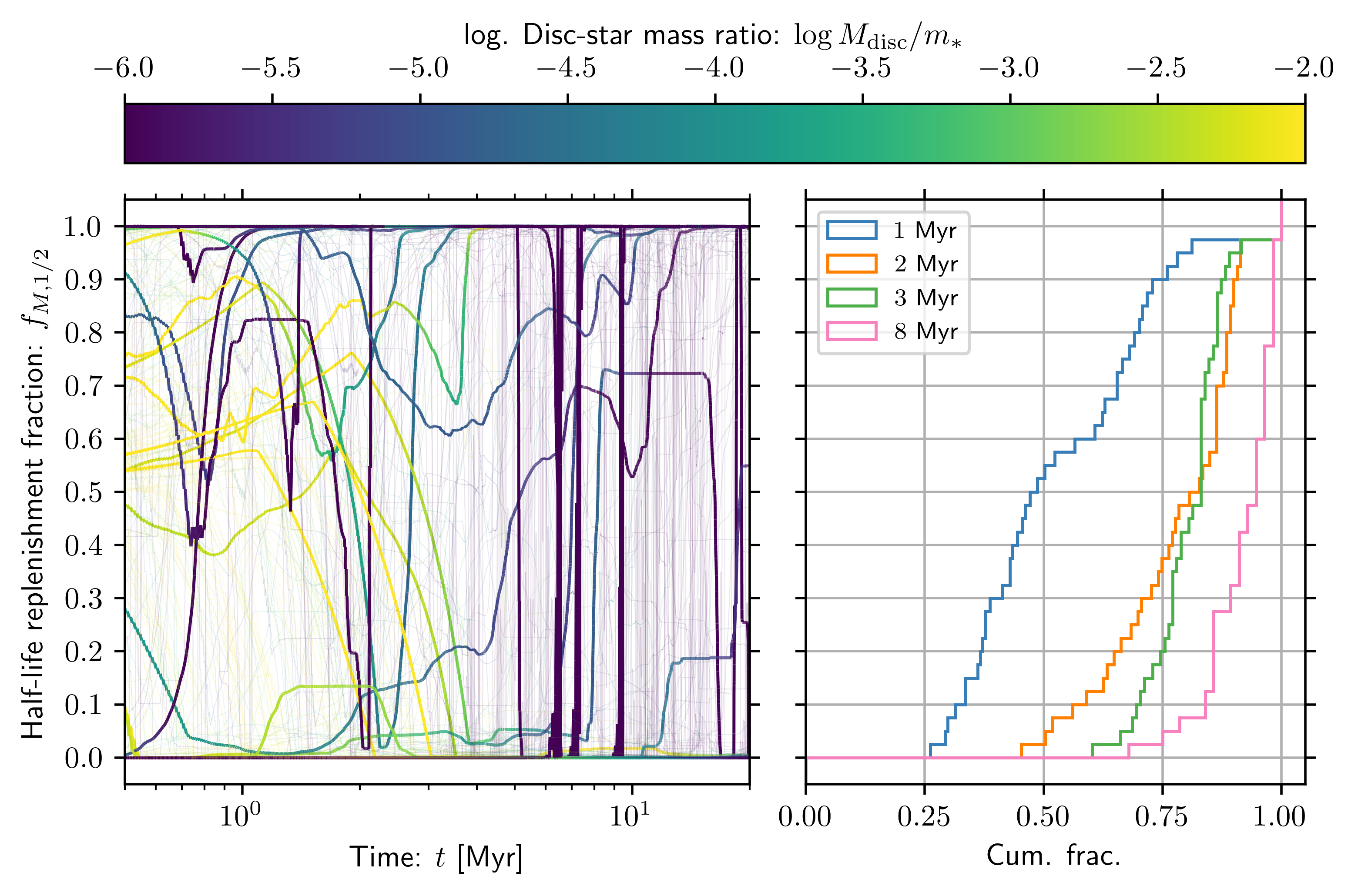}
    \caption{As in Figure~\ref{fig:hlrf}, but including a prescription for the suppression of the BHL accretion by internally-driven photoevaporative winds.}
    \label{fig:hlrf__wind}
\end{figure}

The disc properties we obtain when including photoevaporative winds are shown in Figure~\ref{fig:discprops_wind}. When compared to Figure~\ref{fig:discprops}, these are not strongly influenced by the wind. Early and rapid phases of accretion are not strongly influenced by the wind, and therefore the resultant masses for surviving discs (which have typically undergone faster BHL accretion) are also similar to the negligible wind case. However, we can ask whether these discs are composed of less recently accreted material, since stars are more likely to inhabit regions of low ambient density later in their evolution. Indeed, from Figure~\ref{fig:hlrf__wind} (compared to Figure~\ref{fig:hlrf}), we see a clear decrease in the fraction of replenished material. In particular, for older stars ($\gtrsim 8$~Myr), only $\sim 10$~percent of discs have accreted more than half of their mass recently. Nonetheless, based on our simple estimate, a substantial fraction of discs ($\sim 20-60$~percent at ages $\sim 1{-}3$~Myr) have undergone recent replenishment when photoevaporative winds are considered.

\section{Sensitivity to the lower mass limit of star forming regions }
\label{app:lower_sfrmass_limit}
\restartappendixnumbering

\begin{figure}
    \centering
    \includegraphics[width=0.8\linewidth]{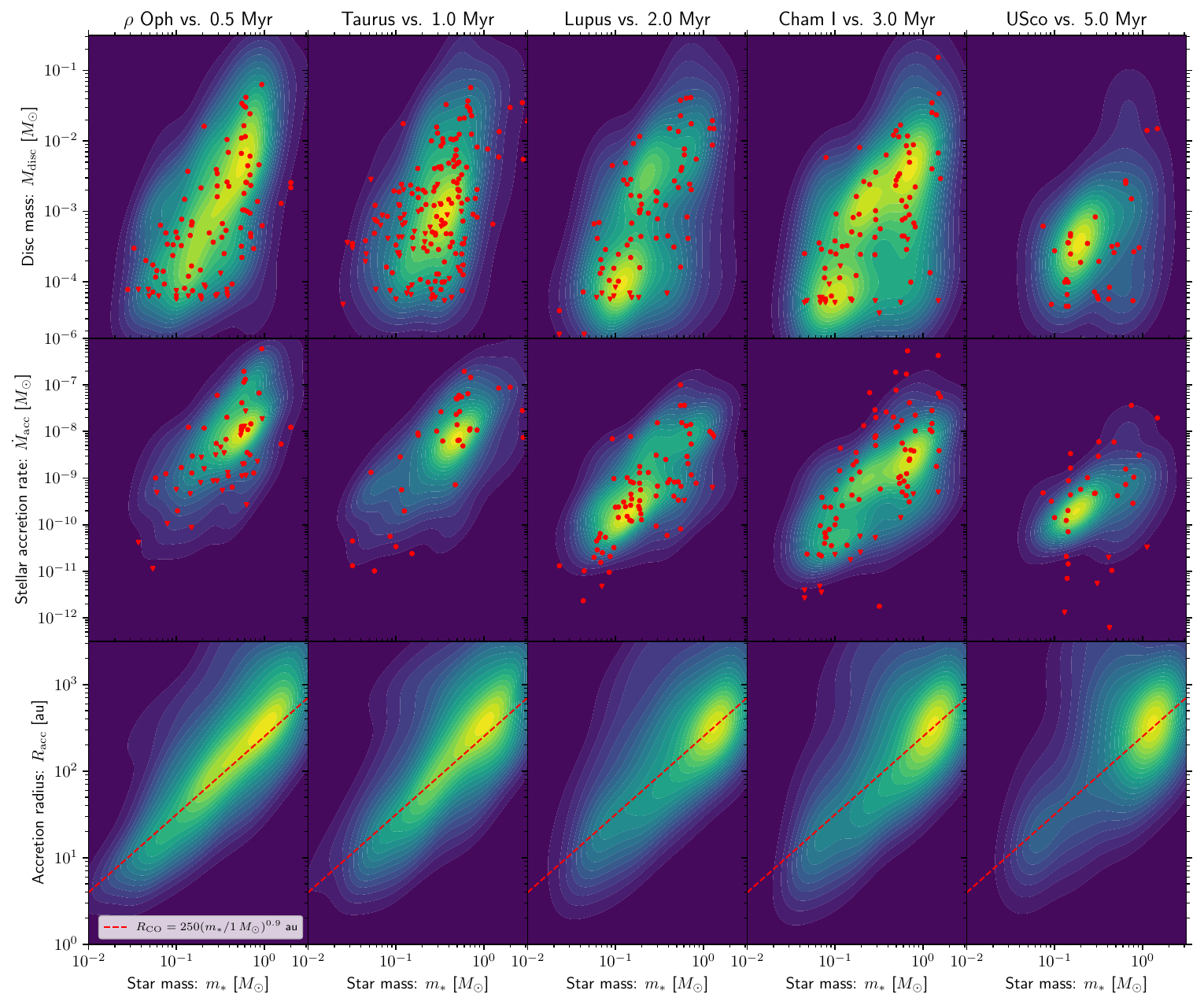}
    \caption{As in Figure~\ref{fig:discprops}, but with a lower mass limit of the collapsing star forming region of $2 \, M_\odot$.}
    \label{fig:discprops_ll1}
\end{figure}

Lower mass star forming regions collapse faster to form stars at a higher ambient density, over a shorter period. We can therefore ask how sensitive our results are to our choice to truncate masses of collapsing star forming regions below $10\,M_\odot$. We therefore repeat the disc evolution experiment, but with a lower cut-off of $2\,M_\odot$ in the masses of star forming region, below which there is not sufficient mass to produce a solar mass star at a star formation efficiency of $50$~percent. 

We show the outcome of our calculations in terms of the disc property distribution in Figure~\ref{fig:discprops_ll1} and the replenishment fraction~\ref{fig:hlrf_ll1}. The distribution of properties in the model remain similar to the fiducial case (Figure~\ref{fig:discprops}). The replenishment fraction drops somewhat for stars of age $\sim 2$~Myr, which can be understood in that fewer stars are still embedded in the bound cloud at intermediate ages. However, overall, the fraction of discs with a a substantial quantity of replenished material remains comparable to our fiducial model. We conclude that our findings are not strongly sensitive on our choice of minimum mass of the star forming regions.

\begin{figure}
    \centering
    \includegraphics[width=0.8\linewidth]{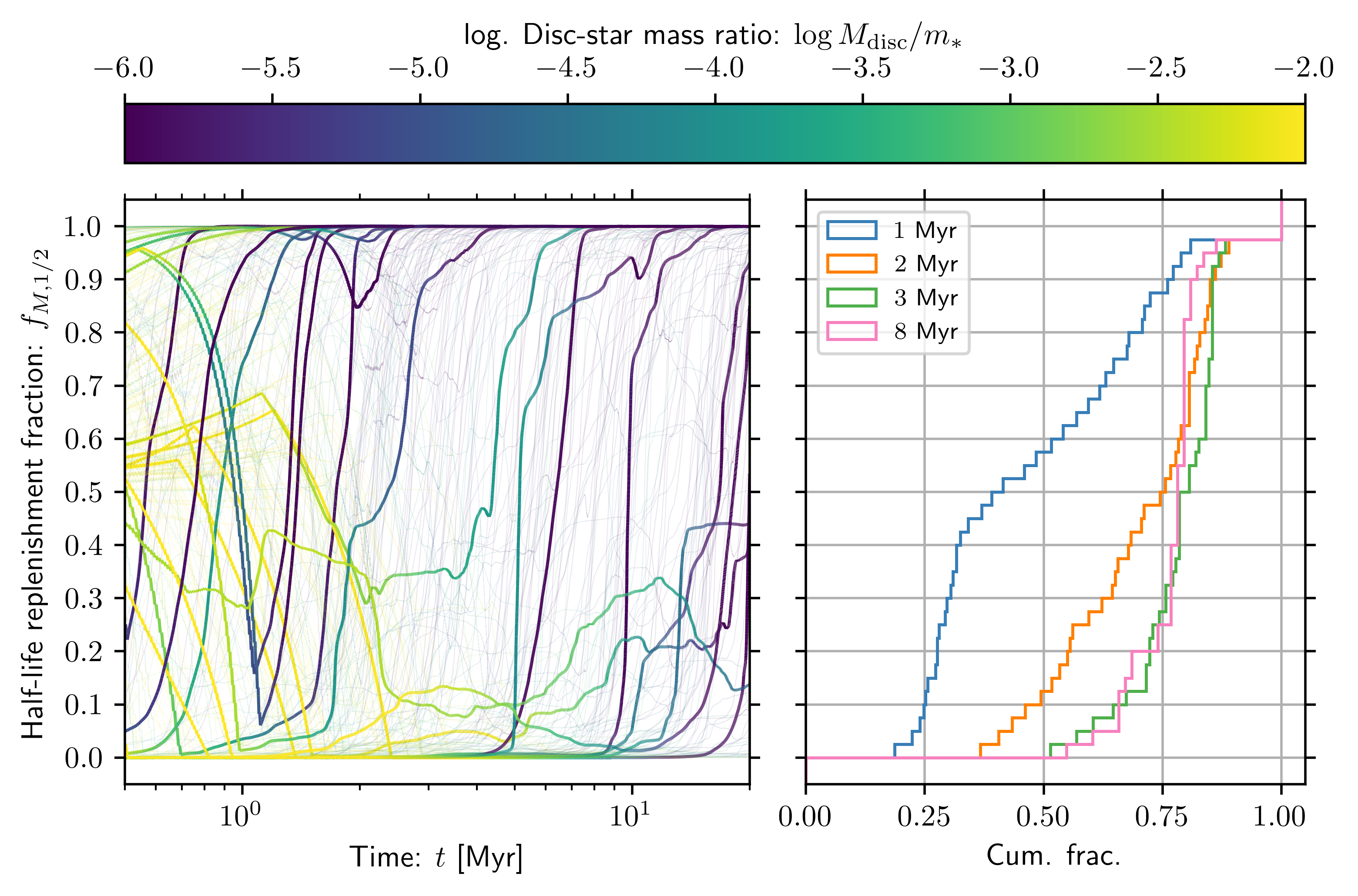}
    \caption{As in Figure~\ref{fig:hlrf}, but with a lower mass limit of the collapsing star forming region of $2 \, M_\odot$.}
    \label{fig:hlrf_ll1}
\end{figure}

\section{Caveats for the BHL accretion rate}

\restartappendixnumbering

\label{sec:other_accretion}
\label{app:other_accretion}
In this work, we have considered a simple estimate for the BHL accretion rate. Here we discuss some possible reasons why the rate may deviate from this approximation, and discuss some open questions about the process of capture of interstellar material.

\subsection{BHL accretion mediated by a disc} 
The nature of BHL accretion has a long history of testing with numerical experiments. In a supersonic, uniform medium, these broadly validate the analytic estimate for the BHL accretion rate within a factor of order unity \citep{Hunt71, Hunt79, Ruffert94, Ruffert96, Edgar04, Mellah15}. In a turbulent medium, the instantaneous accretion rate may be limited by the local vorticity instead of the velocity \citep{Krumholz06}. In this case there may be a substantial reduction in the accretion rate, but only when the Mach number is low (i.e. $\Delta v \sim c_\mathrm{s}$), which is rarely the case in our numerical calculations. In a magnetised medium, the BHL accretion rate may also be reduced by a factor $\sim 2$, although this is dependent on the ratio of magnetic to thermal pressure \citep{Lee14, Burleigh17}. Overall, we expect our simple estimates to yield a reasonable approximation for the mass accretion rate.

\subsection{Cloudlet capture}

Another possible avenue for the capture of interstellar gas is the capture of small cloudlets via tidal energy dissipation. {In pure BHL accretion, the self-gravity of the ISM is ignored, but particularly early in the disc evolution, in a dense environment, the gravitational force exerted by the local medium may be substantial.} In this case, during the close passage between a star and a cloud, the orbital energy may be dissipated by non-radial oscillations within the cloud, analogous to the formation of tight binaries in globular clusters \citep{Rob68, Fabian75}, in a process that remains similar across a range of internal density profiles \citep{Lee86}.  
{Assuming identical physics in the star-cloud to the star-star cases,} capture requires a periastron $a_\mathrm{p}$ during an encounter between a cloud with radius $R_\mathrm{cloud}$ and mass $M_\mathrm{cloud}$ and star with mass $m_*$ such that {\citep{Fabian75}}:
\be
\label{eq:capt_rad}
 {a_\mathrm{p}} \lesssim {a_\mathrm{capt}} \approx {R_\mathrm{cloud}}\left[ \frac{Gm_*}{R_\mathrm{cloud} v_\infty^2}  q_\mathrm{cloud}^{-1}(1+q_\mathrm{cloud}) \right]^{1/6}.
\ee Here $q_\mathrm{cloud}=M_\mathrm{cloud}/m_*$, and $v_\infty$ is the relative velocity between star and cloud at infinite separation. Then  if $a_\mathrm{capt}>R_\mathrm{cloud}$, the cross section is:
\be
    \sigma_\mathrm{capt} \approx \frac{2\pi Gm_*(1+q_\mathrm{cloud}) \left(a_\mathrm{capt} -R_\mathrm{cloud} \right) }{v_\infty^2 }. 
\ee We have made the conservative assumption that only gravitationally focused encounters may result in capture. We exclude encounters for which the star passes through the cloud, which would result in BHL accretion, although even in this case tidal effects may be relevant for a dense ISM. 

The cloud must be compact enough to be captured fully by the star. {We can estimate that the cloud can no longer be captured when the differential potential across the cloud is comparable to the kinetic energy that is dissipated during capture. It is easy to show that this yields $R_\mathrm{cloud} \sim R_\mathrm{BHL}$. Therefore we can estimate the total rate of capture for a cloud of radius $R_\mathrm{cloud}$ as:}
\be
\label{eq:gamma_capt}
    \Gamma_\mathrm{capt}(R_\mathrm{cloud} | \rho_\mathrm{cloud}, \sigma_v, ) = \int_0^{v_{\mathrm{BHL}}} n_\mathrm{clouds}(R_\mathrm{cloud}) \sigma_\mathrm{capt} \tilde{v} g_3(\tilde{v}| \sigma_v) \,\mathrm{d} \tilde{v},
\ee where 
\be
g_{N_\mathrm{D}}(\Delta v | \sigma_v) =  \frac{2 \Delta v^{N_\mathrm{D}-1}}{\Gamma(N_\mathrm{D}/2)} \left(\frac{1}{4\sigma_{v} ^2}\right)^{{N_\mathrm{D}}/2} \exp\left( \frac{-\Delta v^2}{4 \sigma_v^2}\right)
\ee is the Maxwell-Boltzmann distributions  in $N_\mathrm{D}$ dimensions. We have cast equation~\ref{eq:gamma_capt} in terms of the cloud density $\rho_\mathrm{cloud}=  3M_\mathrm{cloud}/ 4 \pi R_\mathrm{cloud}^3$. If large scale turbulence dominates density fluctuations, we can estimate that a typical cloud density will be $\rho_\mathrm{cloud} \sim \rho_\mathrm{gas}$. We have introduced $n_\mathrm{clouds} = 1/(2R_\mathrm{cloud})^3$, where $2R_\mathrm{cloud}$ is the wavelength of the turbulent oscillations in the ISM that produce a cloud of radius $R_\mathrm{cloud}$. Finally, we have defined the BHL velocity $v_{\mathrm{BHL}}$, which is the velocity that yields a particular $R_\mathrm{BHL}=R_\mathrm{cloud}$ when substituted into equation~\ref{eq:BHL_radius}. The overall accretion rate is then $ \dot{M}_{\mathrm{capt}} = \Gamma_\mathrm{capt} M_\mathrm{cloud}$.

\begin{figure}
    \centering
    \includegraphics[width=0.7\columnwidth]{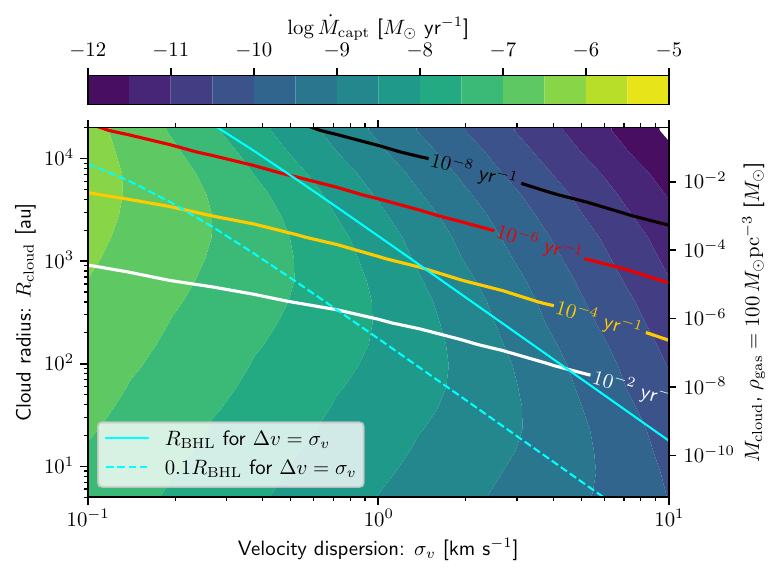}
    \caption{\label{fig:cc_params}The expected mass accretion rate $\dot{M}_\mathrm{capt}$ onto a disc around a star of mass $m_*=1\, M_\odot$ due to gravitationally focused cloudlet capture in a gaseous medium of density $\rho = 100 \, M_\odot $~pc$^{-3} = 6.77 \times 10^{-21}$~g~cm$^{-3}$. The colour bar shows the averaged mass accretion rate, and the rate $\Gamma_\mathrm{capt}$ of individual cloudlet capture is shown by contours. The BHL radius is shown as a solid cyan line, while the optimum clouds contributing to mass accretion have $R_\mathrm{cloud} \approx 0.1 R_\mathrm{BHL}$ (dashed cyan line).}
\end{figure}

In Figure~\ref{fig:cc_params}, we show the results of these calculations for a solar mass star in an ISM of density $\rho_\mathrm{gas} = 100 \, M_\odot $~pc$^{-3}$. We find that capture rates are typically maximised for cloud sizes $R_\mathrm{cloud}\sim 0.1 {-}1 R_\mathrm{BHL}$. Assuming $R_\mathrm{acc} = R_\mathrm{BHL}$ and substituting in $\Delta v_\mathrm{gas} = 1$~km~s$^{-1}$ for a solar mass star into equation~\ref{eq:mdot_BHL} yields $\dot{M}_\mathrm{BHL} \approx 2.4 \times 10^{-8} \,M_\odot$~yr$^{-1}$. Inferring the comparable cloud capture accretion rate from Figure~\ref{fig:cc_params} yields a very similar rate. In general, we expect that tidal cloud capture may play a significant role in disc assembly, alongside BHL accretion. This may increase the rate of interstellar gas capture by a factor of order unity, and may be the subject of future numerical experiment.

\subsection{Face-on accretion} 
\label{sec:sweep-up}
A third mechanism for ISM accretion is face-on disc accretion, where an existing disc produces a large physical cross-section \citep{Moeckel09, Wijnen17}. In this case, the ram pressure between a pre-existing disc and the ISM mediates the accretion process. Face-on accretion may be significant when a star moves through a relatively high velocity medium. In this case, the disc shrinks as a result of deposition of low angular momentum material onto the disc surface \citep{Wijnen17}. There is no steady state for this process unless material can be constantly replenished. This replenishment of outer disc material may be viscous, but then the quantity of material accreted is limited by the initial disc mass. Replenishment via BHL or cloud capture would keep the disc at a radius $R_\mathrm{disc}  \sim R_\mathrm{BHL}$, and would therefore result in a similar accretion rate to BHL accretion \citep{Moeckel09}. However, this may significantly enhance accretion rates when the BHL radius decreases; then $R_\mathrm{disc} > R_\mathrm{BHL}$. Face-on accretion would then temporarily dominate the overall rate, shrinking the disc to maintain a balance between shrinkage and replenishment. Overall, we may expect some moderate enhancement to the rate of capture and stellar accretion during short phases, although this would not be expected to greatly change the overall disc evolution.

\end{document}